\documentclass[aip,amsmath,amssymb,reprint]{revtex4-1} 

\usepackage{graphicx}
\usepackage{latexsym}
\usepackage{bm}

\usepackage{hyperref} 
\usepackage{enumerate}  

\usepackage{stmaryrd} 
\usepackage{upgreek} 
\usepackage[mathscr]{euscript} 

\newcommand{\muSI}{\upmu_0}

\newcommand{\trnsp}{{\!\rm T}} 

\renewcommand{\vec}[1]{\bm{\mathrm{#1}}} 
\newcommand{\vsf}[1]{\bm{\mathsf{#1}}} 
\newcommand{\jump}[1]{\left\llbracket  #1 \right\rrbracket} 

\renewcommand{\d}{d} 
\renewcommand{\ij}{i} 
\newcommand{\Ident}{\textsf{\textbf{I}}} 

\newcommand{\grad}{ \mbox{\boldmath\(\nabla\)}}

\newcommand{\divv}{ \mbox{\boldmath\(\nabla\cdot\)}}
\newcommand{\curl}{ \mbox{\boldmath\(\nabla\times\)}}
\newcommand{\D}{\displaystyle}
\newcommand{\esub}[1]{{\bf e}_{#1}}
\newcommand{\esup}[1]{{\bf e}^{#1}}
\newcommand{\dotv}{  \mbox{\boldmath\(\cdot\)} }
\newcommand{\ddotv}{  {{\textbf :}} }
\newcommand{\cross}{  \mbox{\boldmath\(\times\)} }
\newcommand{\sgn} {\mathrm{sgn}\,}

\newcommand{\const}{{\mathrm{const}}}
\renewcommand{\eqref}[1]{Eq.~(\ref{#1})}

\newcommand{\etal}{\emph{et al.}\ }

\newcommand{\comp}{\,{\scriptstyle{\circ}}\,} 

   \newcommand{\homvOmega}{h^{\bm{\omega}{\scriptsize{\!\times}}\!\vec{v}}_\Omega}
   
  \newcommand{\SRxOmega}{\mathscr{S}_\Omega^{\rm Rx}}
    \newcommand{\Sph}{\SRxOmega} 
    \newcommand{\Sphprime}{\mathscr{S}_{{\rm ph}\Omega'}^{\rm Rx}}

\newcommand{\XHelmult}{\nu_\Omega}

\begin{document}
\title{Extended version, with equation workings: Time-dependent relaxed magnetohydrodynamics -- inclusion of cross helicity constraint using phase-space action} 

\author{R.~L. Dewar}\email{robert.dewar@anu.edu.au}
	\affiliation{Mathematical Sciences Institute, The Australian National University, Canberra, ACT 2601, Australia}
\author{J.~W. Burby},\email{maruchanil1@gmail.com}
	\affiliation{Los Alamos National Laboratory, Los Alamos NM 87545, USA}
\author{Z.~S. Qu}\email{zhisong.qu@anu.edu.au}
	\affiliation{Mathematical Sciences Institute, The Australian National University, Canberra, ACT 2601, Australia}
\author{N. Sato}\email{sato@ppl.k.u-tokyo.ac.jp}
	\affiliation{Graduate School of Frontier Sciences, The University of Tokyo, Kashiwa, Chiba 277-8561, Japan}
\author{M.~J. Hole}\email{matthew.hole@anu.edu.au}
	\affiliation{Mathematical Sciences Institute, The Australian National University, Canberra, ACT 2601, Australia}

\date{\today}
\begin{abstract}
A phase-space version of the ideal MHD Lagrangian is derived from first principles and shown to give a relabeling transformation when a cross-helicity constraint is added in Hamilton's Action Principle. A new formulation of time-dependent \emph{Relaxed} Magnetohydrodynamics (RxMHD) is derived using microscopic conservation of mass, and macroscopic constraints on total magnetic helicity, cross helicity and entropy under variations of density, pressure, fluid velocity, and magnetic vector potential.  This gives Euler--Lagrange equations consistent with previous work on both ideal and relaxed MHD equilibria with flow, but generalizes the relaxation concept from statics to dynamics. The application of the new dynamical formalism is illustrated for short-wavelength linear waves, and the interface connection conditions for Multiregion Relaxed MHD (MRxMHD) are derived. The issue of whether $\vec{E} + \vec{u}\cross\vec{B} = 0$ should be a constraint is discussed. 
\end{abstract}

\maketitle

\section{Introduction}\label{sec:Intro1}

\subsection{Context and Motivation}\label{sec:Motive}

In this paper we are principally concerned with developing RxMHD, a new nondissipative fluid dynamics intermediate between ideal magnetohydrodynamics (IMHD) and relaxed (Rx) magnetohydrostatics (RxMHS), within a single, topologically toroidal domain $\Omega$ that is closed, of genus at least 1, and whose boundary $\partial\Omega$ is smooth, gapless, and perfectly conducting. 

This is part of a larger project, the development of a truly \emph{dynamical} Multiregion Relaxed MHD (MRxMHD); the \emph{static}, MRxMHS version already being well developed and embodied in the SPEC equilibrium code for nonaxisymmetric plasmas. \cite{Hudson_etal_2012b,Dewar_Yoshida_Bhattacharjee_Hudson_2015} The first application of the current dynamicization project is to extend SPEC to include stratified equilibrium flows, \cite{Qu_etal_2020} with codes to model time-dependent behavior to follow later. The SPEC code is already coming to be used in such challenging practical applications as stellarator design, but its main relevance to the present paper is that it is designed to adhere to the principle that a good code should be based on a mathematically well-posed model.

It was pointed out by Grad, \cite{Grad_67} that the problem of constructing nonaxisymmetric toroidal equilibria with nontrivial, smooth pressure profiles using IMHD is \emph{ill-posed} due to singular behavior at resonances on magnetic surfaces with rational magnetic-field-line rotation numbers \emph{and} nonzero pressure gradients. 

The MRxMHD approach has evolved from the mathematical construction of a solution to Grad's problem by Bruno and Laurence, \cite{Bruno_Laurence_96} who showed the existence (sufficiently close to axisymmetry) of weak solutions of the IMHD equations with piecewise constant, stepped pressure profiles. Their construction drew on insights from nonlinear Hamiltonian dynamics, notably KAM theory, to avoid the conjunction of pressure gradients and rational surfaces. This was done by restricting ($\delta$-function) pressure gradients to invariant tori with sufficiently irrational rotation numbers, while having only zero pressure gradients on all irrational surfaces. 

The SPEC code may be viewed \emph{either} as a variational numerical method for finding such weak solutions far from axisymmetry, \emph{or} as a multiregion extension of Taylor's \cite{Taylor_86} plasma relaxation theory, which invokes small-scale turbulence to break nearly all the infinity of IMHD invariants so pressure gradients relax to zero while conserving magnetic helicity. While we use the term ``relaxation'' in this paper, and make some speculative comments about turbulence, we are essentially adopting the first, weak-IMHD view as there is no dissipation in Hamiltonian dynamics. For the purposes of this paper, the term \emph{relaxation} is interpreted as ``relaxation of constraints.''

In the MRxMHD context, $\Omega$ is but a subregion of a larger plasma region, partitioned into multiple relaxation domains physically separated by weak-IMHD current-sheet \emph{interfaces} of zero width. Thus, in general, the boundary $\partial\Omega(t)$ is the union of the inward-facing sides of the interfaces $\Omega$ shares with its neighbors. 

We consider these interfaces to be impervious to magnetic flux, implying the \emph{tangentiality condition}
\begin{equation}\label{eq:tangential}
	\vec{n}\dotv\vec{B} = 0 \:\: \text{on}\:  \partial\Omega \;,
\end{equation}
where $\vec{B} \equiv \curl\vec{A}$ is the magnetic field and $\vec{n}$ is a unit normal at each point on $\partial\Omega$. Also, to conserve magnetic fluxes trapped within $\Omega$, loop integrals of the vector potential $\vec{A}$ within the interfaces must be conserved. \cite{Dewar_Yoshida_Bhattacharjee_Hudson_2015}

We take the interfaces to be perfectly flexible, and impervious to mass and heat transport. However they transmit pressure forces between the subregions so we shall also analyse the interaction between two neighboring regions, $\Omega$ and $\Omega'$. 

This paper carries on the project, started in 2015, \cite{Dewar_Yoshida_Bhattacharjee_Hudson_2015} of ``dynamicizing'' (soft c) MRxMHS. \cite{Hudson_etal_2012b}  This will allow the modeling of low-frequency global modes, linear, unstable, or nonlinearly saturated (limit cycles), as coupled surface waves on the interfaces. However, as these interfaces are infinitely thin, what provides the inertia that determines their finite frequencies?

Clearly, as with other surface waves, the inertia comes from the reaction of the disturbed ambient fluid, which means we can no longer use the purely static relaxation theory used in Ref.~\onlinecite{Hudson_etal_2012b}, which also did not allow for equilibrium flow. However, since the primary role of the fluid dynamics within the ``relaxation regions'' is to endow the interfaces with inertia, one may hope that the global dynamics is insensitive to the detailed mesoscale dynamics within $\Omega$ as long as it is quasi-adiabatic. What is needed is a plasma fluid model that combines the simplicity and well-posedness of MRxMHS with the ability to describe time-dependent flows. It is also desirable to be consistent with low-frequency IMHD where it is applicable, at least in the limit of an infinite number of interfaces

Thus in Ref.~\onlinecite {Dewar_Yoshida_Bhattacharjee_Hudson_2015} we proposed a natural formal extension of the static relaxation theory used in Ref.~\onlinecite{Hudson_etal_2012b}. The 2015 approach led to the ``relaxed plasma'' in $\Omega$ being modeled as an Euler fluid, with the only coupling to the magnetic field occurring at the interfaces. While the Euler fluid model succeeds in endowing the interfaces with inertia, \cite{Dewar_Tuen_Hole_2017} an Euler fluid is very different from an MHD fluid. The present paper makes the fluid model, RxMHD, slightly closer to IMHD by adding a cross-helicity constraint to couple fluid and magnetic field. 

This paper includes sufficient validation tests to be confident that our RxMHD formulation is likely adequate for the purpose outlined above, but more work remains to be done to apply it in toroidal plasma confinement calculations, and also to determine if it has wider theoretical significance and physical application. In particular, it would be interesting to investigate connections with turbulence theories---dynamo effects, inverse cascades, and selective decay---but these topics are outside the scope of the present paper.


\subsection{Lagrangian and Eulerian plasma fluid dynamics}\label{sec:HamPrelim}

In this paper we first recall standard textbook (e.g. Ref.~\onlinecite{Goldstein_80}) classical mechanics, in which one starts with a full \emph{configuration space} of generalized coordinates $q_i$, some of which, say the $q^{\rm holo}_j$, may be subject to \emph{holonomic constraints}, meaning their variations $\delta q^{\rm holo}_j$ are not free but constrained, in that they can be expressed in terms of the remaining, free variations $\delta q^{\rm free}_k$. In the following we use the reduced configuration space spanned by the set of free variables, $q \equiv \left\{q^{\rm free}_k\right\}$, the $\delta q^{\rm holo}_j$ being assumed to be slaved to $q$ by the constraints.

The equations of motion, second-order ordinary differential equations, are derived variationally from a \emph{configuration-space Lagrangian} (CSL) $L(q,\dot{q},t)$ using Hamilton's action Principle $\delta\!\int\! L \,\d t = 0$ to determine which of the possible trial paths through configuration space are true trajectories, where $\delta\! L$ is reduced to a sum over only the $\delta q^{\rm free}_k$ by using the holonomic variational constraints. 

Hamiltonian mechanics halves the order of the equations of motion by doubling the number of free independent variables, configuration space being replaced by the \emph{phase space}, with coordinates comprising both the $q_k$ and the \emph{canonical momenta} $p_k \equiv \partial L/\partial \dot{q}_k$. To make the $p_k$ independent variables, these defining equations reverse roles and are assumed to be solvable for the $\dot{q}_k$ in terms of the $q_k$ and $p_k$ for use in finding the Hamiltonian, $H(q,p,t) \equiv L - \Sigma_k p_k\dot{q}_k$. Then the Hamiltonian equations of motion can be found by first defining the \emph{phase-space Lagrangian} (PSL) $L_{\rm ph}(q,p,t) \equiv \Sigma_k p_k\dot{q}_k - H$,  then applying Hamilton's phase-space action Principle $\delta\!\int\! L_{\rm ph} \,\d t = 0$, $\forall\: \delta p, \delta q$. We provide more on the use and history of the PSL in Subsec.~\ref{sec:Canonical}.

To connect MHD with classical mechanics, we can adopt the \emph{Lagrangian picture} of an MHD fluid as an infinite set of fluid elements, each labeled by its initial position $\vec{x}_0$ and evolving under the Lagrangian time-evolution map, $\vec{x} = \vec{r}_{\vec{v}}^t(\vec{x}_0)$, taking fluid elements from their initial to their current positions. [This map is also called a \emph{flow} in mathematical dynamical systems theory: it is the solution of the dynamical system $\dot{\vec{x}} = \vec{v}(\vec{x},t)$, i.e.
\begin{equation}\label{eq:vFlow}
	\partial_t\vec{r}_{\vec{v}}^t (\vec{x}_0) = \vec{v}(\vec{r}_{\vec{v}}^t,t)\;, \:\: \vec{r}_{\vec{v}}^{t_0}(\vec{x}_0) \equiv \vec{x}_0\:\forall\:\vec{x}_0\:\in\:\Omega_0 \;,
\end{equation}
where $t_0$ is an arbitrary initial time. 
We have added the velocity subscript to indicate which of the two different velocity fields we encounter in the present paper is generating the map.] 

The Lagrangian picture is very useful for providing a physical understanding of fluid dynamics, but its need to attach somewhat arbitrary labels to fluid elements seems to impose an unobservable theoretical construct on the actual physical flow (though, in neutral fluids, particle image velocimetry does make it experimentally possible to visualize Lagrangian trajectories over short times). 

In practice indeed, one normally adopts the \emph{Eulerian picture}, solving PDEs (partial differential equations) for the physically observable fields, which are functions of position $\vec{x}$ and \emph{current} time $t$. In fact it is possible to develop Hamiltonian fluid mechanics in a purely Eulerian way, and we adopt this Eulerian approach in the following, except where it is useful to invoke the Lagrangian picture for enhancing physical understanding.

For maximum accessibility, we develop the presentation using tools already made familiar to fluid and plasma dynamicists in the 1960s; in MHD the seminal groundwork was done by Frieman and Rotenberg \cite{Frieman_Rotenberg_60} and Newcomb. \cite{Newcomb_62} This was generalized to include wave degrees of freedom by Dewar. \cite{Dewar_70} In the latter, the need to develop a theory that included both the holonomically constrained degrees of freedom of the basic MHD fields, mass density, pressure and magnetic field, and the freely variable wave fields, \cite{Whitham_65} was found to be most concisely presented in terms of a general Lagrangian that we have adapted for use in the present paper (Subsec.~\ref {sec:GenCSL}), though for a different purpose.

There has been much important research since the '60s on geometric mechanics, connecting fluid mechanics and MHD with modern mathematics, including differential geometry, some of which is summarized in the recent monograph by Webb. \cite{Webb_18} Of note is the work of Holm \etal \cite{Holm_Marsden_Ratiu_Weinstein_85,Holm_Marsden_Ratiu_98} on the ``Euler--Poincar\'e'' formalism---the history and mathematical ramifications of the duality between the Eulerian and Lagrangian pictures of fluid mechanics and MHD sketched above.

However modern mathematical sophistication is unnecessary for the purposes of the present paper, its avoidance being helped by using the phase-space Lagrangian approach. In this paper we have refrained from using unnecessarily mathematical terminology in order to make the paper accessible to a wider physical-sciences audience. A more mathematical paper may well be necessary for future development of our dynamicization project. 

The plasma is modeled as a magnetohydrodynamic (MHD) fluid. Thus we start by considering the well-known ideal magnetohydrodynamic equations over $\Omega$, which are encapsulated in the Lagrangian, \cite{Newcomb_62,Dewar_70}
\begin{equation}\label{eq:IMHDLag}
	L_\Omega[\vec v, \rho, p, \vec{A}] \equiv \int_\Omega \frac{\rho v^2}{2}\,\d V - W_\Omega \;, 
\end{equation}
with potential energy
\begin{equation}\label{eq:WMHD}
	W_\Omega[p, \vec{A}] \equiv \int_\Omega \left(\frac{p}{\gamma-1} + \frac{B^2}{2\muSI}\right)\d V \;,
\end{equation}
where $\d V$ is the volume element $\d^3 x$ and $[\vec v, \rho, p, \vec{A}]$ signals that $L_\Omega $ is a functional of the Eulerian fields $\vec{v}(\vec{x},t)$, $\rho(\vec{x},t)$, $p(\vec{x},t)$ and $\vec{A} (\vec{x},t) $ --- the fluid velocity, mass density, pressure and magnetic vector potential, respectively (the constant $\muSI$ being the vacuum permeability constant used in SI units). 

We shall later verify that the IMHD equation of motion can be derived from this Lagrangian by defining the \emph{action integral}
\begin{equation}\label{eq:Actiondef}
	\mathscr{S} \equiv \int\!\!  L_\Omega\, \d t \;, 
\end{equation}
and deriving an Euler--Lagrange equation from Hamilton's Principle (of stationary action), $\delta \mathscr{S} = 0$.

\subsection{Outline of paper}\label{sec:Outline}

In Sec.~\ref{sec:Intro2} we first review the IMHD equations and define the infinite-dimensional configuration space of the CSL. In Subsec.~\ref{sec:IMHDmicro},we review the microscopic holonomic constraints of IMHD and give an elementary interpretation of them as a Lie symmetry. In Subsec.~\ref{sec:FIhel}, we list IMHD macroscopic global invariants, and in Subsec.~\ref{sec:Relax} we explain their relation to the RxMHD concept: in summary, we modify ideal dynamics by relaxing the continua of local, holonomic constraints on $p$ and $\vec{B}$, replacing this infinity of constraints with a finite set of global constraints to conserve a few IMHD global invariants, which is our \emph{definition} of relaxation.

In Sec.~\ref{sec:GenCSL} we present a representation for a general CSL (see Subsec.~\ref{sec:HamPrelim}) for fluids which allows for arbitrary arrays of both holonomically constrained and free fields (the continuum analogs of the $q^{\rm holo}$ and $q^{\rm free}$ generalized coordinates above), thus forming an appropriately general starting point for developing our relaxation formalism. This allows the integrations by parts to derive general Euler--Lagrange equations from Hamilton's Principle to be reused in different scenarios, rather than redoing the integrations by parts each time. 

As an example of the use of this formalism, in Subsec.~\ref{sec:IMHDCSL} we derive the standard equation of motion in momentum-conservation form, using the IMHD Lagrangian. However, we also show that adding to this Lagrangian a cross-helicity constraint term, which should be redundant as it is preserved under IMHD time evolution, gives physically \emph{incorrect} Euler--Lagrange equations. This unsatisfactory property of the CSL is the main motivation for developing the PSL approach as a successful alternative for constructing our variational relaxation theory.

In Sec.~\ref{sec:GenPSL} we derive and motivate the PSL (see Subsec.~\ref{sec:HamPrelim}) approach: In Subsec.~\ref{sec:Canonical} we derive the IMHD Hamiltonian as defined on the phase space $\vec{x},\bm{\pi}$, where $\bm{\pi}$ is a canonical momentum density. In Subsec.~\ref{sec:StdForm}, we then make a change of variable $\bm{\pi} = \rho\vec{u}$, the Hamiltonian now being defined on the \emph{non}canonical, $\vec{x},\vec{u}$ phase space. 

In the PSL approach momenta are varied freely, so $\vec{u}$ is not constrained to be $\dot{\vec{x}} = \vec{v}$. Nevertheless, when used with the IMHD Hamiltonian, we show $\vec{u}$ is indeed the correct Eulerian velocity field. Modifying the IMHD Hamiltonian with a ``redundant'' cross-helicity constraint term we find, unlike with the CSL, the constrained PSL still gives $\vec{u}$ as the physically correct flow velocity. However, the cross-helicity constraint does break the identification of $\vec{u}$ with $\vec{v}$, relegating $\vec{v}$ to the role of generating fluid-element labels advecting on a reference flow.
   
In Sec.~\ref{sec:vRxph} we dynamicize equilibrium relaxation theory by taking as Hamiltonian the relaxed-MHD-equilibrium energy functional \cite{Finn_Antonsen_83} to form a PSL.  In Subsec.~\ref{sec:RxMHDEL} this is used in the phase-space version of Hamilton's Principle to give dynamical Euler--Lagrange equations, which are analyzed in Subsec.~\ref{sec:RxMHDuB}. 

As a preliminary investigation of the physical implications of our newly derived dynamics, in Sec.~\ref{sec:WKBRxMHD} we derive the local dispersion relations for linear waves in the WKB approximation for both IMHD and RxMHD and find them very different. In the RxMHD case, at least some waves break the ideal Ohm's Law. 

In Sec.~\ref{sec:IOhm} we discuss whether and how to make the ideal Ohm's Law a constraint. Conclusions are given in Sec.~\ref{sec:Concl}.
 
In Appendix~\ref{sec:genTaylor} we review the derivation of fully relaxed plasma equilibria, with field-aligned flow, by finding stationary points of an energy functional that includes the helicity and cross helicity constraints (our RxMHD Hamiltonian), and in Appendix~\ref{sec:FAcomparison} we show that the PSL approach allows a natural extension to axisymmetric 
equilibria with cross-field flow.\cite{Finn_Antonsen_83, Hameiri_83, Dennis_Hudson_Dewar_Hole_14a} The Grad--Shafranov-Bernoulli equations for such equilibria are given in Sec.~\ref{sec:GSeq}. 

Finally, in Appendix~\ref{sec:RxMHDELsurf}, the coupling across the interfaces between two neighboring relaxation regions is derived variationally from the phase-space action principle and shown to be the standard pressure-jump condition found previously. \cite{Dewar_Yoshida_Bhattacharjee_Hudson_2015} Thus generalization to MRxMHD is straightforward.

\section{Ideal MHD (IMHD) constraints and invariants and the relaxation concept}\label{sec:Intro2}

\subsection{Conservation constraint PDEs, equation of motion, and configuration space}\label{sec:IMHD}

In this subsection we give the evolution equations for the four fields $\{\rho,p,\vec{B},\vec{v}\}$ defining the \emph{state} of the system at any given time In each evolution equation we first give the \emph{conservation} version and then the equivalent \emph{advective} form, which is in terms of the \emph{advective} derivative $\d /\d t \equiv \partial_t + \vec{v}\dotv\grad$, the total derivative along Lagrangian fluid-element trajectories (paths) $\vec{r}^t(\vec{x}_0)$.

Microscopic (fluid-element-wise) conservation of mass is expressed in the continuity equation 
\begin{equation}\label{eq:Continuity}
	\partial_t\rho + \divv(\rho\vec{v}) = 0 \:\: \Leftrightarrow \:\: \frac{\d\rho}{\d t}  = -\rho\divv\vec{v} \;,
\end{equation}
and microscopic entropy conservation in the ideal adiabatic pressure equation
\begin{equation}\label{eq:AdiabaticPressure}
	\partial_t p + \divv(p\vec{v}) + (\gamma - 1)p\divv\vec{v} = 0\:\: \Leftrightarrow \:\: \frac{\d p}{\d t}  = -\gamma p\divv\vec{v} \;.
\end{equation}

The ``freezing in'' of magnetic flux into microscopic loops, advected by the flow field $\vec{v}$, \cite{Newcomb_58} is expressed by
\begin{equation}\label{eq:FrozenFlux}
	\partial_t \vec{B} - \curl(\vec{v}\cross\vec{B}) = 0\:\: \Leftrightarrow \:\: 
	\frac{\d \vec{B}}{\d t}  = -\vec{B}\dotv(\Ident\divv\vec{v} - \grad \vec{v}) \;,
\end{equation}
where $\vsf{I}$ is the unit dyadic. This equation can be derived by eliminating $\vec{E}$ from the ``ideal Ohm's Law,'' $\vec{E} + \vec{v}\cross\vec{B} = 0$ by taking the curl of both sides and using the ``pre-Maxwell'' form of Faraday's law $\curl\vec{E} = -\partial_t\vec{B}$.

We shall refer to the above three equations as the \emph{IMHD constraint PDEs} as they represent microscopic constraints on the time evolution of the set of the IMHD fluid attributes $\{\rho,p,\vec{B}\}$ along path lines.

In the Eulerian picture, we define an \emph{evolution} of the state of an MHD system as the set of functions $\{\rho,p,\vec{B},\vec{v}\}$ over some interval of $t$. As we are developing a variational method, we regard these as trial-function evolutions, which at this point need obey neither the constraint PDEs nor the \emph{equation of motion},
\begin{equation}\label{eq:idealEqMot}
	\rho\frac{\d\vec{v}}{\d t} = -\grad p + \frac{1}{\muSI}(\curl \vec{B})\cross\vec{B} \;.
\end{equation}

We define the infinite-dimensional MHD \emph{configuration space} as the space on which $\vec{v}$ is defined (or, more generally, to which $\vec{v}$ is tangent). \emph{Feasible} MHD evolutions solve the holonomic constraint PDEs [in the case of IMHD, Eqs.~(\ref{eq:Continuity}--\ref{eq:FrozenFlux})], slaving the constrained members of the set $\{\rho,p,\vec{B}\}$ to $\vec{v}$. \emph{Autonomous} MHD evolutions are not only feasible, but also solve \eqref{eq:idealEqMot} so do not require external forcing.

\subsection{Microscopic variational constraints and Lie symmetry}\label{sec:IMHDmicro}

In variational IMHD the fields $\rho$, $p$, and $\vec{B}$ are not free variables but are constrained holonomically to evolve under the same Lagrangian map as the fluid elements.  The variation generator $\bm{\xi}$ is defined on the same configuration space as $\vec{v}$, which is kinematically constrained to vary with $\bm{\xi}$ in the first of the four IMHD constraint equations in Eulerian form below
\begin{align}
	\delta\vec{v} &= \partial_t\bm{\xi} + \vec{v}\dotv\grad\bm{\xi} - \bm{\xi}\dotv\grad\vec{v}\;, \label{eq:vvar}\\
	\delta\rho &= -\divv(\rho\,\bm{\xi})\;, \label{eq:rhovar}\\
	\delta p &= -\gamma p\divv\bm{\xi} - \bm{\xi}\dotv\grad p\;, \label{eq:pvar}\\
	 \delta\vec{B} &= \curl(\bm{\xi}\cross\vec{B}) \nonumber\\
	 	&= -\vec{B}\divv\bm{\xi} + \vec{B}\dotv\grad\bm{\xi} - \bm{\xi}\dotv\grad\vec{B}
		\label{eq:Bvar}\;. 
\end{align}
These are Eqs. (4.6--4.9) of Newcomb,\cite{Newcomb_62} his $\bm{\varepsilon}$ being our $\bm{\xi}$, a more common notation. 

The holonomically constrained Eulerian variations were also discussed and used in Ref.~\onlinecite{Dewar_Yoshida_Bhattacharjee_Hudson_2015}, the precursor of the current paper. The difference between Eulerian ($\delta$) variations, and their corresponding Lagrangian ($\Delta$)  variations, connected via the operator equation $\Delta \equiv \delta + \bm{\xi}\dotv\grad$, \cite{Dewar_70} was also reviewed there---by definition $\delta\vec{x} \equiv 0$, so applying the operator $\Delta$ to $\vec{x}$ gives $\Delta\vec{x} = \bm{\xi}$. 

In the Lagrangian picture, \cite{Frieman_Rotenberg_60} $\bm{\xi}(\vec{x},t)$ is the variation $\delta\vec{r}_{\vec{v}}^t (\vec{x}_0)$ at fixed $\vec{x}_0$ and $t$, but with $\vec{x}_0$ expressed in terms of $\vec{x}$ and $t$ by inverting the Lagrangian map, i.e. $\bm{\xi}(\vec{x},t) \equiv \delta\vec{r}_{\vec{v}}^t \comp(\vec{r}_{\vec{v}}^t)^{-1}(\vec{x},t)$.  The variations of $\rho, p, $\vec{B}$, $ (\ref{eq:rhovar}--\ref{eq:Bvar}) can be derived by integrating the Lagrangian versions of Eqs.~(\ref{eq:Continuity}--\ref{eq:FrozenFlux}) along \emph{varied} Lagrangian trajectories. 

As a consistency check of the Eulerian holonomic variations Eqs.~(\ref{eq:vvar}--\ref{eq:Bvar}), not reliant on the Lagrangian picture, we show in Appendix~\ref{sec:Lie} that the constraint PDEs, Eqs.~(\ref{eq:Continuity}--\ref{eq:FrozenFlux}), are preserved under perturbation by $\bm{\xi}$. This shows $\bm{\xi}$ generates a Lie symmetry (i.e. an infinitesimal transformation taking solutions to solutions) in the $\{\rho,p,\vec{B}\}$ subspace of the state space. We now argue this implies all feasible configurations are continuously (diffeomorphically) connected by transformations generated by all differentiable functions $\bm{\xi}$. 

 To construct finite transformations from the infinitesimal generator $\bm{\xi}$, label functions of $\vec{x},t$ arbitrarily with a configuration evolution parameter, $\tau$ say. E.g., $\bm{\xi}$ becomes $\bm{\xi}(\vec{x},t,\tau)$, $\vec{v}$ becomes $\vec{v}(\vec{x},t,\tau)$ etc. By interpreting $\delta$ in Eqs.~(\ref{eq:rhovar}--\ref{eq:Bvar}) as the operator $\partial_\tau$ (and $\Delta$ as $\partial_\tau + \bm{\xi}\dotv\grad$) we evolve $\rho$, $p$, and $\vec{B}$ away from any given solution of the constraint equations at $\tau=0$ into a continuously connected family of feasible evolutions.  

However, not all such evolutions are physical, as the MHD equation of motion, \eqref{eq:idealEqMot},
is \emph{not} automatically preserved under perturbation by $\bm{\xi}$ using Eqs.~(\ref{eq:rhovar}--\ref{eq:Bvar}). [Rather, for the perturbed equation of motion to be satisfied, $\bm{\xi}$ must satisfy the \emph{linearized} equation of motion, see e.g. Eq.~(25) of Ref.~\onlinecite{Frieman_Rotenberg_60}].

Along any given curve through IMHD evolution space, the action, \eqref{eq:Actiondef}, is $\mathscr{S}(\tau)$ and Hamilton's Principle becomes the requirement that, for a evolution to be autonomous, it must be such that $\mathscr{S}'(\tau) = 0$ at that evolution on all feasible evolution families passing through it (i.e. for all $\bm{\xi}$ along the given evolution.
It will be verified in Sec.~\ref{sec:IMHDCSL} that the equation of motion can be derived from Hamilton's Principle using the holonomic variations above. 

In Sec.~\ref {sec:RxMHDPSL} we develop our formal relaxation procedure as a reduction in the number of state variables constrained to vary with $\bm{\xi}$ from the four fields $\{\rho,p,\vec{B},\vec{v}\}$ to two, $\{\rho,\vec{v}\}$. That is, in RxMHD, $p$ and $\vec{B}$ are no longer restricted to a feasible set, instead being treated as freely variable fields (constrained only globally by Lagrange multiplier terms added to the Lagrangian).

\subsection{Macroscopic (global) IMHD invariants}\label{sec:FIhel}

There is an infinity of microscopic IMHD invariants applying within infinitesimal fluid elements and tubes, but the only macroscopic (i.e. global within a domain $\Omega$) IMHD invariants we shall use as nonholonomic constraints are
\begin{itemize}
\item the \emph{magnetic helicity} $2\muSI K_\Omega $, where, \cite{Bhattacharjee_Dewar_82} we define the invariant $K_\Omega $ as
\begin{equation}\label{eq:Helicity}
	K_\Omega [\vec{A}] \equiv \frac{1}{2\muSI}\int_\Omega  \vec{A}\dotv\vec{B} \, \d V  
\end{equation}
\item the cross helicity $\muSI K_\Omega^{\rm X}$, where
\begin{equation}\label{eq:XHel}
	K_\Omega^{\rm X} [\vec{v},\vec{A}] \equiv \frac{1}{\muSI}\int_\Omega\!\vec{v}\dotv\vec{B}\,\d V 
\end{equation}
(this global invariant derives from a relabelling symmetry in the Lagrangian representation of the fields \cite{Salmon_88,Padhye_Morrison_96a,Padhye_Morrison_96b,Webb_Zank_07,Webb_etal_2014_I,Webb_etal_2014_II,Araki_2015}),
\item the total entropy \cite{Dewar_Yoshida_Bhattacharjee_Hudson_2015}
\begin{equation}\label{eq:Entropy}
	S_\Omega [\rho,p] \equiv
	\int_\Omega \frac{\rho}{\gamma - 1} \ln\left(\kappa\frac{p}{\rho^\gamma}\right) \,\d V \;.
\end{equation}
\end{itemize}

\subsection{The relaxation concept}\label{sec:Relax}

Relaxation of a plasma is often taken to mean an approach to a steady state.  As high-temperature plasmas have very low particle collision rates, collisional dissipative mechanisms of relaxation, like resistivity and viscosity, may be very slow. Instead the physical mechanism for plasma relaxation is usually taken to be some kind of small-scale turbulence (see e.g. Ref.~\onlinecite{Taylor_86}), though in strongly three-dimensional systems deterministic chaos has also been invoked. \cite{Hudson_etal_2012b}
 
Even ignoring the anisotropy created by the strong confining magnetic field, we could perhaps discern the existence of four relaxation timescales, an electromagnetic (or Alfv\'en) timescale $\tau^{\rm EM}_{\rm Rx}$, a thermal equilibration timescale $\tau^{\rm T}_{\rm Rx}$, a turbulent dynamo  \cite{Squire_Bhattacharjee_2016,Moffatt_2014,Yokoi_2013} decay timescale $\tau^{\rm T}_{\rm dyn}$ and an electrostatic potential equilibration timescale.  Except perhaps for the latter two effects (which are relevant to the discussion in Sec.~\ref{sec:IOhm}) we are not concerned with timescales in this paper, but assume simply there is an upper bound, $\tau_{\rm Rx}$, beyond which our relaxation theory becomes applicable. 
 
A serious discussion of the complex physics of relaxation mechanisms is beyond the scope of this paper. Instead we define what we mean by relaxation formalistically, as a generalization of the postulate of Taylor \cite{Taylor_86} that a relaxed steady state can be found by minimizing (``relaxing'') an energy functional, subject to the constraint that only the most macroscopically robust, global invariant of IMHD, the magnetic helicity, survives for $t \gtrsim \tau_{\rm Rx}$. 
This allows the frozen-in flux constraint \cite{Newcomb_58} to be broken so that topological changes in the magnetic-field-line flow can occur when energetically favorable, allowing the formation of magnetic islands and chaotic regions.

While seemingly over simplistic, Taylor's approach was found to be remarkably effective for describing experimental results from a very turbulent toroidal magnetic confinement experiment, Zeta. 

In the MRxMHD equilibrium approach, \cite{Dewar_Yoshida_Bhattacharjee_Hudson_2015} Taylor relaxation has the great attraction that it reduces the problem of computing $\vec{B}$ in $\Omega$ to that of solving a well-studied elliptic PDE (the linear-force-free, or \emph{Beltrami} equation). This solves the long-standing mathematical problem, \cite{Grad_67} of the existence of IMHD equilibria in nonaxisymmetric toroidal plasmas by regularizing away the singularities that arise if the magnetic field lines are constrained to lie on smoothly nested invariant tori (magnetic surfaces). Instead, because the Beltrami equation is elliptic, solving it requires no assumptions as to the detailed behavior of magnetic field lines, so magnetic islands and chaos cause no problems. 

In this paper we explore the question: Can the MRxMHD approach be extended to slowly time-dependent problems and equilibria with flow?

\begin{figure}[htbp]
\centering
\includegraphics[width=8.5cm]{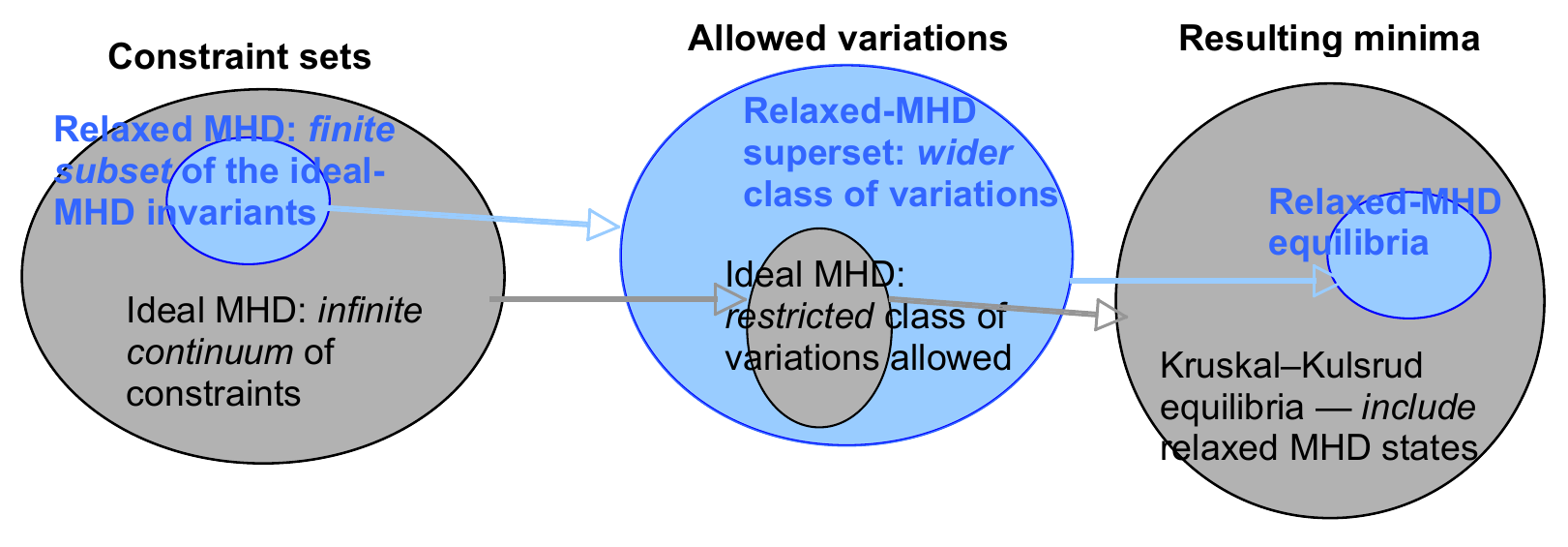}
\caption{Constraint sets, spaces of allowed variations, and equilibrium states:  Illustrating how reducing the number of constraints, broadening the space of allowed variations, narrows the class of equilibria, and \emph{vice versa}. (Reprinted with permission from \emph{Entropy}. \cite{Dewar_etal_2008})}
\label{fig:Venn}
\end{figure}

We follow Taylor in asssuming that most of the microscopic invariants of IMHD are broken even in less turbulent systems, allowing heat transport and magnetic and vorticity reconnection events that allow the system to evolve to a self-organized, relaxed steady state. However, we increase the number of IMHD global invariants used as constraints in energy minimization so as to widen the class of energy minima (see FIG.~\ref{fig:Venn} for a Venn diagram).

Only a negligible strength of the ideal-invariant-breaking mechanism should be needed to maintain such a steady state, once formed, in a near-collisionless plasma. Thus we take as an ``axiom'' that the \emph{steady state} Euler--Lagrange equations from a variational relaxation principle \cite{Dewar_etal_2008} should be consistent with the original ideal equations. That is, for a mathematical formulation of relaxation to be physically acceptable it should satisfy the Principle of Consistency with Ideal Equilibria (\emph{Consistency Principle} for short): \emph{Relaxed equilibria should be a subset of the stationary solutions of the IMHD equations}. (A problem with this principle is discussed in Sec.~\ref{sec:IOhm}.) 
 
To achieve relaxation we remove the microscopic holonomic constraints from $p$ and $\vec{B}$, replacing them with the three macroscopic IMHD constraints in Sec.~\ref {sec:FIhel}. These include the fluid-magnetic cross helicity \eqref{eq:XHel}, which couples an unconstrained plasma flow $\vec{u}$ and vector potential $\vec{A}$, thus improving on an earlier attempt at deriving RxMHD, \cite{Dewar_Yoshida_Bhattacharjee_Hudson_2015} (though at the expense of complicating the PDE for the magnetic field).  

Note that we have not lifted the holonomic constraint on $\rho$ as, for fluid elements to have any physical meaning, it would seem we need the mass density not only to be defined (perhaps in a weak, coarse-grained sense) but to obey a mass continuity equation pointwise, which we enforce variationally using \eqref{eq:rhovar}, $\delta\rho = -\divv(\rho\,\bm{\xi})$ (so we do not need to include total mass as a global constraint).
 
 A relaxation approach for finding equilibria with flow by constraining cross helicity was used by Finn and Antonsen \cite{Finn_Antonsen_83} using an \emph{entropy maximization} relaxation principle 
 (see also the contemporaneous paper by Hameiri \cite{Hameiri_83}). However, they show this leads to the same equations as energy minimization. Thus we take, as in IMHD, the entropy in $\Omega$ to be conserved and follow Taylor in defining relaxed states as \emph{energy minima}. 
  
 Pseudo-dynamical energy-descent relaxation processes that conserve topological invariants have been developed, \cite{Vallis_Carnevale_Young_89,Vladimirov_Moffatt_Ilin_99} but we stay within the framework of conservative classical mechanics by developing a dynamical formalism, RxMHD, that includes relaxed equilibria as stationary points of a relaxation Hamiltonian, with Lagrange multipliers to constrain chosen macroscopic invariants, but also allows non-equilibrium motions. Stability can also be examined by taking the second variation of the Hamiltonian, \cite{Vladimirov_Moffatt_Ilin_99} but in this paper we deal only with first variations.
 
\section{General Configuration-Space Lagrangian (CSL)}\label{sec:GenCSL}

\subsection{General Lagrangian}\label{sec:HolonomicFreeSplit}

As in Ref.~\onlinecite{Dewar_70}, consider a configuration-space Lagrangian of the general form
\begin{equation}\label{eq:Lgen}
	L[\vec{r},\vec{v},\bm{\upeta}] \equiv \int_\Omega\mathcal{L}(\vec{v},\bm{\uplambda},\bm{\upeta},\partial_t\bm{\upeta},\grad\bm{\upeta}) \,\d V \;,
\end{equation}
where $\bm{\upeta}$ is any set of \emph{freely variable}, unconstrained physical fields (scalars and 3-vectors, 
but not including $\bm{\xi}$), such as $p$ and $\vec{A}$ in the case of RxMHD, arranged into a matrix column vector. The set $\bm{\uplambda}$ is similarly comprised of some physical variables, such as $\rho$, $p$ and $\vec{B}$ in the case of IMHD, that are functions of $\vec{x}$ and $t$, but are \emph{holonomically constrained} to vary with  
$\bm{\xi}$, as discussed in Sec.~\ref{sec:IMHDmicro}. 

Explicit examples of $\bm{\uplambda}$, $\bm{\upeta}$, and $\mathcal{L}$ are given at the end of this section and in other sections of the paper, but it is worthwhile here to elaborate a little further on the IMHD examples above---in this case the transpose $\bm{\uplambda}^\trnsp$ is the row vector $[\rho,p,\vec{B}]$, its holonomic constraint equation [an instance of \eqref{eq:deltalambda} below] summarizing Eqs.~(\ref{eq:rhovar}--\ref{eq:Bvar}). 

On the other hand $\bm{\upeta}$ is made up of any other fields \emph{not} subject to these constraints---in Ref.~\onlinecite{Dewar_70}, $\bm{\upeta}$ contained wave variables (amplitudes and phases) but in this paper $\bm{\upeta}$ comes into play when we relax holonomic constraints. For instance, in Taylor's relaxation principle the only local constraint on $\vec{B}$ is $\divv\vec{B} = 0$, which is enforced by the representation $\vec{B} = \curl\vec{A}$. So $\bm{\uplambda} = \bm{0}$ and the vector potential $\vec{A}$, being a free variable, appears in $\bm{\upeta}$. 
Note also that another difference between Ref.~\onlinecite{Dewar_70} and the present paper is that here $\bm{\xi}$ denotes the variation in the Lagrangian time evolution map, \eqref{eq:vFlow}, $\Delta\vec{x}(\vec{x},t) = \delta\vec{r}_{\vec{v}}^t (\vec{x}_0)$, whereas in Ref.~\onlinecite{Dewar_70} $\bm{\xi}$ represented waves on a Lagrangian mean flow.

In the assumed absence of an external potential (e.g. gravity) in the system, the $\vec{x}$ and $t$ dependences of $\mathcal{L}$ arise only from those of the physical fields $\vec{v}$,  $\bm{\uplambda}$ and $\bm{\upeta}$, and their derivatives. 

Using a somewhat more explicit version of the formalism used in Ref.~\onlinecite{Dewar_70}, we can represent some or all of the holonomic Eulerian variations in Eqs.~(\ref{eq:rhovar}--\ref{eq:Bvar}) in the general form
\begin{equation}\label{eq:deltalambda}
	\delta\bm{\uplambda}^\trnsp =\bm{\uplambda}^\trnsp\dotv(\vsf{V}\dotv\grad\bm{\xi} - \vsf{\Lambda}\divv\bm{\xi}) - \bm{\xi}\dotv{\grad}\bm{\uplambda}^\trnsp \;,
\end{equation}
where the diagonal, dimensionless constraint \emph{structure matrices} $\vsf{V}$ and $\vsf{\Lambda}$ have as elements real-numbers, zero 3-vectors, and symmetric dyadic-tensor elements occurring only on their diagonals. (In $\vsf{V}$ all nondyadics are zero---its role is to project out the 3-vector component of $\bm{\uplambda}^\trnsp$.) 

The dot product $\bm{\cdot}$ denotes the usual 3-vector inner product, and also a matrix product where appropriate. (If there is no $\bm{\cdot}$ between 3-vectors then they form a dyadic). The transpose operation ${}^\trnsp\!$ acts on both matrices and dyadics, e.g. $(\vec{a}\vec{b})^\trnsp = \vec{b}\vec{a}$. Also, the real number $\divv\bm{\xi}$ distributes multiplicatively over the elements of the matrix $\vsf{\Lambda}$ in the standard way, and dotting with the dyadic $\grad\bm{\xi}$ likewise distributes over the elements of $\vsf{V}$, with the convention that a product of a zero element and a dyadic remains a null element of unchanged type.  

\subsection{General CSL Euler--Lagrange equations}\label{sec:GenEL}

In the following, $\delta L/\delta f$ represents the standard \emph{functional derivative} of $L$ with respect to an arbitrary field $f$, see e.g. the review by Morrison. \cite{Morrison_98}
In Sec.~\ref{sec:IMHDmicro} we showed that the Eulerian constraint variations could be verified without any appeal to the Lagrangian picture. It is also well known \cite{Newcomb_62,Dewar_70} that Hamilton's Principle can be applied using only the current coordinates $\vec{x}$ and  Eulerian variations $\delta$. 
Thus, for variations of compact support localized in time and space so that boundary terms can be omitted, the variation in the action integral \eqref{eq:Actiondef} is
\begin{align}
	\delta \mathscr{S} 
	&= \!\iint\!\!\left[\delta\vec{v}\dotv\frac{\delta \mathcal{L}}{\delta\vec{v}} 
	+ \delta\bm{\uplambda}^\trnsp \dotv\frac{\delta\mathcal{L}}{\delta\bm{\uplambda}} 
	+ \delta\bm{\upeta}^\trnsp \dotv \frac{\delta\mathcal{L}}{\delta\bm{\upeta}} \right]\d V\d t \nonumber\\ %
	&= \!\iint\!\!\left[
	(\partial_t\bm{\xi} + \vec{v}\dotv\grad\bm{\xi} - \bm{\xi}\dotv\grad\vec{v})
								\dotv\frac{\partial\mathcal{L}}{\partial\vec{v}} 
		+ \delta\bm{\upeta}^\trnsp \dotv\frac{\delta\mathcal{L}}{\delta\bm{\upeta}} \right. \nonumber\\
	&\qquad\left. + \left[\bm{\uplambda}^\trnsp\! \dotv(\vsf{V}\dotv\grad\bm{\xi} - \vsf{\Lambda}\divv\bm{\xi}) 
					- \bm{\xi}\dotv\grad\bm{\uplambda}^\trnsp\right] \!\dotv
	\frac{\partial\mathcal{L}}{\partial\bm{\uplambda}} \right]\d V\d t \nonumber\\ %
	&= \!\iint\!\!\left\{
					\bm{\xi}\dotv\!\left[ - \partial_t\left(\frac{\partial \mathcal{L}}{\partial\vec{v}}\right) 
					- \divv\left(\vec{v}\frac{\partial \mathcal{L}}{\partial\vec{v}} \right)
	 				\right.\right.\nonumber\\
	&
	\qquad\qquad\left.\left.\mbox{} - \grad\vec{v}\dotv\frac{\partial \mathcal{L}}{\partial\vec{v}} 
	+ \frac{\delta L}{\delta\vec{r}}\right] + \delta\bm{\upeta}^\trnsp \dotv\frac{\delta\mathcal{L}}{\delta\bm{\upeta}}
	\right\}\d V\d t \label{eq:Lgenvar}
	\;,
\end{align}
with
\begin{align}
	\frac{\delta L}{\delta\vec{r}} = 
	\divv\!\left(\vsf{I}\, \bm{\uplambda}^\trnsp \!\dotv\vsf{\Lambda}\dotv\frac{\partial\mathcal{L}}{\partial\bm{\uplambda}}
	-\vsf{V}\dotv\bm{\uplambda}^\trnsp \frac{\partial\mathcal{L}}{\partial\bm{\uplambda}}  
		\right)
	- (\grad\bm{\uplambda}^\trnsp)\dotv\frac{\partial\mathcal{L}}{\partial \bm{\uplambda}} \label{eq:deltaLbydeltar}\;,
\end{align}
using the assumed symmetry of its dyadic blocks to commute $\vsf{V}$ with $\bm{\uplambda}^\trnsp$. 

In the above equations, $\delta L/\delta\vec{r}$ represents that part of the scalar-product coefficient of $\bm{\xi}$ on the RHS of \eqref{eq:Lgenvar} (i.e. the terms in square brackets []) not contributed by the three terms in $\delta\vec{v}$, \eqref{eq:vvar}. The $\delta\vec{r}$ in the denominator of $\delta L/\delta\vec{r}$ is a simplification of $\delta\vec{r}_{\vec{v}}^t (\vec{x}_0) \equiv \bm{\xi}(\vec{x},t)$ (see Subsec.~\ref{sec:HolonomicFreeSplit}). Referring to the summary of classical mechanics in Sec.~\ref{sec:Intro1}, $\delta L/\delta\vec{r}$ is the analog of $\partial L/\partial q$.

Hamilton's Principle, $\delta\mathscr{S} = 0 \: \forall \: \bm{\xi}$, now gives the Euler--Lagrange equation
\begin{align}
	&\partial_t\left(\frac{\partial \mathcal{L}}{\partial\vec{v}}\right) 
		+\divv\left(\vec{v}\frac{\partial \mathcal{L}}{\partial\vec{v}} \right)
	 	+\grad\vec{v}\dotv\frac{\partial \mathcal{L}}{\partial\vec{v}} = \frac{\delta L}{\delta\vec{r}} \;, \label{eq:rEL} 
\end{align}

Using \eqref{eq:deltaLbydeltar} the equation of motion \eqref{eq:etaEL} can be put in partial conservation form, cf. Ref.~\onlinecite[Eq.~(24)]{Dewar_70},
\begin{align}
	&\partial_t\!\left(\frac{\partial \mathcal{L}}{\partial\vec{v}}\right) 
	+ \divv\!\left[\vec{v}\frac{\partial \mathcal{L}}{\partial\vec{v}} 
	+ \vsf{V}\dotv\bm{\uplambda}^\trnsp \frac{\partial\mathcal{L}}{\partial\bm{\uplambda}}  
	+ \vsf{I}\left(\mathcal{L} - \bm{\uplambda}^\trnsp\! \dotv\vsf{\Lambda}\dotv\frac{\partial\mathcal{L}}{\partial\bm{\uplambda}}\right)\right] 
	\nonumber\\
	&\qquad= \grad\mathcal{L} - (\grad\vec{v})\dotv\frac{\partial \mathcal{L}}{\partial\vec{v}} 
	- (\grad\bm{\uplambda}^{\!\rm T})\dotv\frac{\partial\mathcal{L}}{\partial \bm{\uplambda}}
	\;, \label{eq:ELrcons}
\end{align}
where $\divv(\Ident\mathcal{L}) \equiv \grad\mathcal{L}$ has been added to both sides of \eqref{eq:ELrcons} so that, in the absence of an external potential, the RHS is the remaining part of $\grad\mathcal{L}$ obtained by applying the chain rule to all the arguments of $\mathcal{L}$ except $\vec{v}$ and $\bm{\lambda}$. That is [see \eqref{eq:Lgen}] RHS $= (\grad\bm{\upeta}^\trnsp)\dotv\partial\mathcal{L}/\partial\bm{\upeta} + (\grad\bm{\upeta}_t^\trnsp)\dotv\partial\mathcal{L}/\partial\bm{\upeta}_t + \grad[(\grad\bm{\upeta})^\trnsp]\dotv\partial\mathcal{L}/\partial(\grad\bm{\upeta})$, where the transpose ${}^\trnsp$ in the last term turns the column vector $\grad\bm{\upeta}$ into a row vector containing the transposes of any dyadics in $\grad\bm{\upeta}$. (This ensures that  ``$\bm{\upeta}$ contracts with an $\bm{\upeta}$, $\grad$ contracts wth a $\grad$.'')

When $\mathcal{L}$ depends only on $f$ and not on $\partial_t f$ or $\grad f$, then $\delta L/\delta f = \partial \mathcal{L}/\partial f$, which identity has been used extensively to simplify \eqref{eq:Lgenvar}. However $\mathcal{L}$ does not depend so simply on $\bm{\upeta}$ --- instead we have, on integration by parts with respect to $t$ and $\vec{x}$,
\begin{equation}\label{eq:deltaSbyeta}
	\frac{\delta\mathscr{S}}{\delta\bm{\upeta}} = \frac{\partial \mathcal{L}}{\partial\bm{\upeta}} 
	-  \partial_t\frac{\partial \mathcal{L}}{\partial\bm{\upeta}_t} - \divv\frac{\partial \mathcal{L}}{\partial\grad\bm{\upeta}} \;,
\end{equation}
where $\bm{\upeta}_t$ denotes $\partial_t \bm{\upeta}$ and $\partial\mathcal{L}/\partial\vsf{\upeta}$ and $\partial\mathcal{L}/\partial\grad\vsf{\upeta}$ are column vectors of derivatives of $\mathcal{L}$ with respect to the elements of $\vsf{\upeta}$ and the gradients of these elements, respectively.

Using \eqref{eq:deltaSbyeta}, the free-field Euler--Lagrange equations follow from Hamilton's Principle, $\delta\mathscr{S}/\delta\bm{\upeta} = 0$,  
\begin{align}
	\partial_t\frac{\partial \mathcal{L}}{\partial\bm{\upeta}_t} + \divv\frac{\partial \mathcal{L}}{\partial\grad\bm{\upeta}} &= \frac{\partial \mathcal{L}}{\partial\bm{\upeta}} \;. \label{eq:etaEL}
\end{align}

Dotting both sides of \eqref{eq:etaEL} with $\grad\bm{\upeta}^\trnsp$ from the left and subtracting the results from both sides of \eqref{eq:ELrcons}, the full momentum conservation result expected from Noether's Theorem is found to be, cf. Ref.~\onlinecite[Eq.~(27)]{Dewar_70},
\begin{equation}\label{eq:momcon}
	\partial_t\vec{G} + \divv\vsf{T} = \bm{0} \;,
\end{equation}
where
\begin{align}
	\vec{G} &\equiv \frac{\partial \mathcal{L}}{\partial\vec{v}} 
	- (\grad\bm{\upeta}^\trnsp) \dotv\frac{\partial \mathcal{L}}{\partial\bm{\upeta}_t}  
	\label{eq:Gdef} 
\end{align}
and
\begin{align}
	\vsf{T} &\equiv \vec{v}\frac{\partial \mathcal{L}}{\partial\vec{v}} 
	+ \vsf{V}\dotv\bm{\uplambda}^\trnsp \frac{\partial\mathcal{L}}{\partial\bm{\uplambda}}  
	+ \vsf{I}\left(\mathcal{L} - \bm{\uplambda}^\trnsp\! \dotv\vsf{\Lambda}\dotv\frac{\partial\mathcal{L}}{\partial\bm{\uplambda}}\right) 
	\nonumber\\&\qquad
	 - \text{Tr}\left(
	 \frac{\partial\mathcal{L}}{\partial\grad\bm{\upeta}}\grad\bm{\upeta}^\trnsp 
	               \right) \;,
	\label{eq:Tdef}
\end{align}
the trace operator Tr contracting over the indices of $\bm{\upeta}$, but \emph{not} of $\grad$. 
The right-hand side of \eqref{eq:momcon} vanishes because of the cancellation between the RHS of \eqref{eq:ELrcons} and terms arising from the subtraction process. An energy conservation equation can also be derived, as in Ref.~\onlinecite{Dewar_70}.

\subsection{Example: Ideal MHD CSL with cross helicity constraint}\label{sec:IMHDCSL}

As an explicit example, consider an MHD Lagrangian 
\begin{equation}\label{sec:XHelConstrainedCSL}
	L_\Omega[\vec v, \rho, p, \vec{B}] \equiv \int_\Omega\d V \frac{\rho\vec{v}^2}{2}  - W_\Omega + \nu K_\Omega^{\rm X}[\vec{v},\vec{B}] \;,
\end{equation}
with $W_\Omega$ given by \eqref{eq:WMHD} and $\rho, p, \vec{B}$ constrained within each fluid element to conserve mass and entropy, and to ``freeze-in'' magnetic flux. These constraints are expressed in the time evolution equations (\ref{eq:Continuity}--\ref{eq:FrozenFlux}) and the holonomic variations given in Eqs.~(\ref{eq:rhovar}--\ref{eq:Bvar}). 

We have also added a global constraint term, $\nu K_\Omega^{\rm X}[\vec{v},\vec{B}]$, where $\nu$ is a Lagrange multiplier to enforce constancy of $K_\Omega^{\rm X}$, the cross helicity \eqref{eq:XHel}. As the cross helicity is an IMHD invariant, one might expect this constraint to be redundant but we shall find that it actually leads to an \emph{incorrect} equation of motion when $\nu \neq 0$. For the purposes of this paper this is a fatal flaw in the CSL approach.

In the compact representation, \eqref{eq:deltalambda}, the constrained quantities are combined into a matrix column vector $\bm{\uplambda}$, made up of two scalars, $\rho$ and $p$, and a 3-vector, $\vec{B}$,
\begin{equation}\label{eq:MHDlambda}
	\bm{\uplambda}^\trnsp = [\rho,p,\vec{B}] \;.
\end{equation}

Then the Lagrangian density is
\begin{align}
	\mathcal{L} 
	&= \frac{\rho\vec{v}^2}{2} - \frac{p}{\gamma - 1} - \frac{\vec{B}\dotv\vec{B}}{2\muSI} + \nu\frac{\vec{v}\dotv\vec{B}}{\muSI} 
	\nonumber\\
	&= \frac{\uplambda_1\vec{v}^2}{2} - \frac{\uplambda_2}{\gamma - 1} - \frac{\bm{\uplambda_3}\dotv\bm{\uplambda_3}}{2\muSI} + \nu\frac{\vec{v}\dotv\bm{\uplambda_3}}{\muSI} \;, \label{eq:MHDCSL}
\end{align}
which has no free fields $\bm{\upeta}$.

By comparing Eqs.~(\ref{eq:rhovar}--\ref{eq:Bvar}) and \eqref{eq:deltalambda} we see that the structure matrices are
\begin{equation}
	\vsf{V} = \left[\begin{array}{ccc} 0 & 0 & \bm{0} \\0 & 0 & \bm{0} \\ \bm{0} & \bm{0} & \vsf{I}\end{array} \right] \;,
	\quad
	\vsf{\Lambda} = \left[\begin{array}{ccc} 1 & 0 & \bm{0} \\ 0 & \gamma & \bm{0} \\ \bm{0} & \bm{0} & \vsf{I}\end{array}\right] \;,
	\label{eq:strucmats1}
\end{equation}
with $\bm{0}$ denoting the zero 3-vector.

Thus \eqref{eq:Gdef} gives
\begin{align}
	\vec{G} & = \uplambda_1\vec{v} + \nu\frac{\bm{\uplambda_3}}{\muSI} 
	\equiv \rho\vec{v} +\frac{\nu}{\muSI} \vec{B}
	\;,
	\label{eq:GMHD} 
\end{align}
and \eqref{eq:Tdef} gives
\begin{align}
	\vsf{T} &= 
	\lambda_1\vec{v}\vec{v} + \frac{\nu\vec{v}\bm{\uplambda_3}}{\muSI}
	+ \left[0,0,\bm{\uplambda}_3\right] \left[\frac{v^2}{2},\, \frac{1}{\gamma - 1},\, 
	\frac{\nu\vec{v} - \bm{\uplambda}_3}{\muSI}\right]^\trnsp \nonumber\\
	&\quad + \vsf{I}\left(\frac{\uplambda_1\vec{v}^2}{2} - \frac{\uplambda_2}{\gamma - 1} 
		- \frac{\bm{\uplambda_3}\dotv\bm{\uplambda_3}}{2\muSI} + \nu\frac{\vec{v}\dotv\bm{\uplambda_3}}{\muSI} \nonumber\right.\\
	&\qquad\left.\mbox{} - \frac{\uplambda_1\vec{v}^2}{2} + \gamma\frac{\uplambda_2}{\gamma - 1} 
	      - \frac{\bm{\uplambda}_3\dotv(\nu\vec{v} - \bm{\uplambda}_3)}{\muSI} \right)
	\nonumber\\ %
	&= \rho\vec{v}\vec{v}
	+ \vsf{I}\left(p + \frac{\vec{B}\dotv\vec{B}}{2\muSI}\right) - \frac{\vec{B}\vec{B}}{\muSI} \nonumber\\
	&\qquad\quad \mbox{}
	+ \frac{\nu}{\muSI}(\vec{v}\vec{B} + \vec{B}\vec{v})
	\;. \label{eq:TMHD}
\end{align}
When the Lagrange multiplier $\nu = 0$, $\vec{G}$ and $\vsf{T}$ are the standard MHD momentum density and total stress tensor, respectively, thus providing a verification both of the general formalism and of the specific CSL, \eqref{eq:MHDCSL}.  

However, when $\nu \neq 0$, $\vec{G}$ and $\vsf{T}$ have no obvious physical interpretation. If the terms proportional to $\nu$ canceled out in the momentum conservation equation \eqref{eq:momcon}, the constraint would at least have no physical effect. However, writing the equation of motion for $\vec{B}$ in \eqref{eq:FrozenFlux} as $\partial_t\vec{B} + \divv(\vec{v}\vec{B} - \vec{B}\vec{v}) = \bm{0}$ we see that cancellation cannot occur because the contribution of the cross-helicity term to the stress tensor is symmetric rather that antisymmetric. 

A problem with the globally constrained CSL approach was also found previously, \cite{Sato_Dewar_2017} when applied to the Euler flow Lagrangian with fluid helicity as a constraint. This was found not to give a physically correct Bernoulli equation. These examples lead to the conclusion that applying global constraints to a CSL cannot be relied upon to give a physically meaningful model, motivating our development of the PSL as an alternative in the following.

\section{Phase-space Action Principle for General MHD-like fluids}\label{sec:GenPSL}

In developing relaxed MHD (RxMHD) we follow Ref.~\onlinecite{Dewar_Yoshida_Bhattacharjee_Hudson_2015} in maintaining the microscopic (holonomic) IMHD constraint \eqref{eq:rhovar} on variations in mass density $\rho$, intrinsic to the concept of fluid element, so that total mass is automatically conserved under variation.  Also as in Ref.~\onlinecite{Dewar_Yoshida_Bhattacharjee_Hudson_2015} we vary $p$ freely, and $\vec{A}$ freely within $\Omega$ but holonomically constrained on $\partial\Omega$. 
Then the above global invariants are enforced by using Lagrange multipliers, the main departure from Ref.~\onlinecite{Dewar_Yoshida_Bhattacharjee_Hudson_2015} being the inclusion of the cross helicity as a constraint to couple magnetic field and fluid in the relaxation process. 

In this section we develop a general variational principle that uses two velocity fields in representing the motion of the plasma fluid: $\vec{u}(\vec{x}, t)$, defined purely in the Eulerian picture in a given Lab frame, 
and $\vec{v}(\vec{x}, t|\vec{x}_0, t_0)$, the vector field of the dynamical system $\dot{\vec{x}} = \vec{v}$ that provides a Lagrangian labeling of the fluid elements.

As the global invariants of ideal MHD form such an essential part of our relaxation theory we first review them before deriving the phase-space Lagrangian approach and testing it on IMHD in the presence of an imposed (redundant) cross-helicity constraint.

\subsection{Canonical Hamiltonian formulation}\label{sec:Canonical}

Building on the general Lagrangian formulation set out in Sec.~\ref{sec:GenCSL}, we define the \emph{canonical momentum densities}
\begin{align}
	\bm{\pi} &\equiv \frac{\partial \mathcal{L}}{\partial\vec{v}} \;, \label{eq:pidef} \\
	\bm{\uppi}_\upeta &\equiv  \frac{\partial \mathcal{L}}{\partial\bm{\upeta}_t} \label{eq:pietadef} \;,
\end{align}
where $\mathcal{L}$ is a CSL density as in Sec.~\ref{sec:HolonomicFreeSplit}. 
We now suppose these equations to be solved to give $\vec{v}$ 
and $\bm{\upeta}_t$ as functions of $\bm{\pi}$ and $\bm{\uppi}_\upeta$,
with corresponding Hamiltonian
defined by the Legendre transformation
\begin{align}
	H[\vec{r},\bm{\pi},\bm{\upeta},\bm{\uppi}_\upeta,t] &= \int_\Omega \mathcal{H}\,\d V \nonumber\\
	\text{where}\:\: \mathcal{H}(\vec{r},\bm{\pi},\bm{\upeta},\bm{\uppi}_\upeta,t)  &\equiv \bm{\pi}\dotv\vec{v} + {\bm{\uppi}_\upeta}^{\!\trnsp}\dotv\bm{\upeta}_t - \mathcal{L} 
\label{eq:Hgen}\;.
\end{align} 
The general variation of $H$ is
\begin{align}
	\delta H 
	&= \int\!\d V\!\left[  
		\delta\bm{\pi}\dotv\vec{v} 
		+ \left(\bm{\pi} - \frac{\partial \mathcal{L}}{\partial\vec{v}}\right)\!\dotv\delta\vec{v} 
		- \bm{\xi}\dotv\frac{\delta L}{\delta\vec{r}} + \delta{\bm{\uppi}_\upeta}^{\!\trnsp}\bm{\upeta}_t  \right.\nonumber\\
	&\qquad\left.\mbox{} 
		+ \left({\bm{\uppi}_\upeta} - \frac{\partial \mathcal{L}}{\partial\bm{\upeta}_t}\right)^\trnsp\delta\bm{\upeta}_t 
	- \delta\bm{\upeta}^\trnsp\left(\frac{\partial \mathcal{L}}{\partial\bm{\upeta}}
		- \divv\frac{\partial \mathcal{L}}{\partial\grad\bm{\upeta}}\right) 
 \right] \nonumber \\ %
	&=  \int\!\d V\!\left[\,\delta\bm{\pi}\dotv\vec{v} - \bm{\xi}\dotv\frac{\delta L}{\delta\vec{r}} \right.\label{eq:Hgenvar} \\
	&\quad\left.\mbox{} + \delta{\bm{\uppi}_\upeta}^{\!\trnsp}\bm{\upeta}_t 
	- \delta\bm{\upeta}^\trnsp\left(\frac{\partial \mathcal{L}}{\partial\bm{\upeta}} - \divv\frac{\partial \mathcal{L}}{\partial\grad\bm{\upeta}}\right)\right]
		\: \forall\,\delta\vec{v},\, \delta\bm{\upeta}
		\;. \nonumber
\end{align}
where we used \eqref{eq:pidef} and \eqref{eq:pietadef}.
Thus
\begin{align}
	\vec{v} & = \frac{\delta H}{\delta\bm{\pi}} \;, \quad \frac{\delta H}{\delta\vec{r}} = -\frac{\delta L}{\delta\vec{r}} \;, 
	\label{eq:HgenELxipi} \\
	\bm{\upeta} _t & = \frac{\delta H}{\delta\bm{\uppi}_\upeta} \;, \quad \frac{\delta H}{\delta\bm{\upeta}} 
	= -\frac{\partial\mathcal{L}}{\partial\bm{\upeta}} + \divv\frac{\partial \mathcal{L}}{\partial\grad\bm{\upeta}} \;,  \label{eq:HgenELetapieta} 
\end{align}
the first equations of \eqref{eq:HgenELxipi} and \eqref{eq:HgenELetapieta} being the obvious generalizations of the canonical Hamilton equation of motion $\dot{q}_i = \partial H/\partial p_i$. The generalizations of $\dot{p}_i = -\partial H/\partial q_i$, though less standard, are provided by eliminating $\mathcal{L}$ from \eqref{eq:rEL} and \eqref{eq:etaEL} using \eqref{eq:HgenELxipi} and \eqref{eq:HgenELetapieta},
\begin{align}
	&\partial_t\bm{\pi} + \divv\left(\frac{\delta H}{\delta{\bm{\pi}}}\bm{\pi}\! \right) 
	+ \left(\grad\frac{\delta H}{\delta{\bm{\pi}}}\right)\!\dotv\bm{\pi}
	= -\frac{\delta H}{\delta\vec{r}} \;, \label{eq:Hgeneqmotpi}\\
	&\qquad\qquad\qquad \partial_t\bm{\uppi}_\upeta = -\frac{\delta H}{\delta\bm{\upeta}} \label{eq:Hgeneqmoteta}
	\;.
\end{align}

We shall not elaborate on these canonical equations further, as we do not use them in this paper. Instead we build on the concept of the \emph{phase-space Lagrangian} (PSL), 
$L_{\rm ph}\!\left[\vec{r},\vec{v},\bm{\pi},\partial_t\bm{\pi},\bm{\upeta},\bm{\upeta}_t, {\bm{\uppi}_\upeta},\partial_t {\bm{\uppi}_\upeta} \right]$,
\begin{equation}\label{eq:PSL}
	L_{\rm ph} \equiv \int_\Omega\!\!\left(\bm{\pi}\dotv\vec{v} + {\bm{\uppi}_\upeta}^{\!\trnsp} \dotv\bm{\upeta}_t \right)\,\d V - H \;,
\end{equation}
with the corresponding \emph{phase-space action},
\begin{equation}\label{eq:PSA}
	\mathscr{S}_{\rm ph} \equiv \int\!\! L_{\rm ph}\, \d t \;.
\end{equation}

It is a standard result in classical mechanics that a ``modified Hamilton's Principle'' (see e.g. Ref.~\onlinecite[p. 362]{Goldstein_80}), or \emph{phase-space action principle}  
\begin{equation}\label{eq:PSAP}
	\delta\mathscr{S}_{\rm ph} = 0  
\end{equation}
for \emph{all} phase-space variations $\delta q_i$ and $\delta p_i$ ($\bm{\xi}$,  $\delta\bm{\upeta}$, $\delta\bm{\pi}$ and $\delta\bm{\uppi}_{\upeta}$ in our case), yields the  canonical Hamiltonian equations of motion and thus provides a valid alternative to the original configuration-space-based Hamilton's Principle for deriving physical equations of motion. 

Note that, as the $p_i$ ($\bm{\pi}$ and $\bm{\uppi}_{\upeta}$ in our case) are now regarded as \emph{freely variable}, the dimensionality of the space of allowed variations is doubled in the phase-space action principle,  making it much more flexible as the $p_i$ are now untied from their Lagrangian roots in $\partial L/\partial\dot{q}_i$. 

That is, by using the PSL action principle we are no longer restricted to canonical Hamiltonian mechanics as the variational principle remains valid under \emph{noncanonical} changes in phase-space coordinates. By appealing directly to the phase-space action principle, extra formal complications such as noncanonical Poisson brackets\cite{Morrison_98} can be avoided.

Note particularly that our PSL is of the \emph{same} general form as the CSL of Sec.~\ref{sec:GenCSL}, except with the set of free variables $\bm{\upeta}$ augmented by including $\bm{\pi}$ (or its replacement under a change of phase-space variables). Thus, once we have a Hamiltonian,  \emph{we can reuse the general Euler--Lagrange results of Sec.~\ref{sec:GenEL} simply by replacing $\mathcal{L}$ with $\mathcal{L}_{\rm ph}$.}
For such reasons we make the phase-space action principle the basis of the theory developed in this paper.

Historical note: The phase-space action principle has long been used (implicitly) in the generating-function theory of canonical transformations, Ref.~\onlinecite[p. 380]{Goldstein_80}, though the current terminology and emphasis on its utility in noncanonical transformations is more recent (see e.g. Refs.~\onlinecite{Cary_Littlejohn_83, Cary_Brizard_09}). A more mathematical terminology for $L_{\rm ph}\d t$ is the fundamental, \cite{Littlejohn_82} or Poincar\'{e}-Cartan, Ref.~\onlinecite[p. 44]{Arnold_89}, 1-form. 

\subsection{PSL for standard form Lagrangians}\label{sec:StdForm}

Although we do not need the canonical equation of motion, we do need to make explicit the canonical Hamiltonian $H$ in order to form the phase-space Lagrangian \eqref{eq:PSL} as the starting point. This is greatly simplified by restricting, in this paper, to CSLs of the standard kinetic-minus-potential energy form,
\begin{equation}\label{eq:Lstd}
	\mathcal{L}_{\rm std} = \frac{\rho\vec{v}^2}{2} - \mathcal{V}(\bm{\uplambda},\bm{\upeta},\grad\bm{\upeta}) \;,
\end{equation}
where we have assumed $\mathcal{V}$ contains neither $\bm{\upeta}_t$ nor $\vec{v}$ (the fields in $\bm{\upeta}$ are \emph{passive} in the sense to be defined in Sec.~\ref{sec:RxMHDuB}). The former assumption implies $\partial\mathcal{L}/\partial\bm{\upeta}_t = 0$ and the latter implies $\partial\mathcal{L}/\partial\vec{v} = \rho\vec{v}$. Thus, from \eqref{eq:pidef}, we can eliminate \vec{v} in terms of $\bm{\pi}$ trivially, $\vec{v} = \bm{\pi}/\rho\,$. Also, from \eqref{eq:pietadef}, $\bm{\uppi}_\upeta = 0$. Thus, from \eqref{eq:Hgen}, the canonical Hamiltonian density is $\mathcal{H}_{\rm std} = \bm{\pi}^2/2\rho + \mathcal{V}$.

However, in this paper we do \emph{not} work with the canonical momentum $\bm{\pi}$ but instead exploit the freedom afforded by the PSL to work with a velocity-like phase space variable $\vec{u}$ obtained by the \emph{noncanonical} change of variable $\bm{\pi} = \rho\vec{u}$. (This $\vec{u}$, $\vec{v}$ formalism was introduced by Burby. \cite{Burby_2017}) Then the Hamiltonian becomes  
\begin{align}
	H_{\rm nc} 
	&= \int_\Omega\left(\frac{\rho\vec{u}^2}{2} + \mathcal{V}\right)\,\d V
	\label{eq:Hstd}\;,
\end{align}
and the PSL in noncanonical form becomes, from \eqref{eq:PSL}, 
\begin{equation}\label{eq:PSLnc}
	L_{\rm nc} \equiv \int_\Omega\!\!\rho\vec{u}\dotv\vec{v} \,\d V - H_{\rm nc} \;.
\end{equation}

This equation forms the basis of the development in the remainder of this paper. As $\bm{\pi}$ was freely variable using the PSL action principle, so $\vec{u}$ is freely variable in the noncanonical phase space. It has the dimensions of a velocity, and, as we shall show, it can indeed be interpreted as an Eulerian flow velocity, freed from the labeling constraint of the Lagrangian flow velocity $\vec{v}$.\\

\noindent\textbf{Note:} Typically the only field gradient in $\mathcal{V}$ is that of $\vec{A}$, in $\vec{B}=\curl\vec{A}$. As the curl makes it clumsy to work with the general Euler--Lagrange equation \eqref{eq:etaEL} it is useful to give here the functional derivative $\delta L_{\rm nc}/\delta\vec{A}$ when $\mathcal{V}$ is an explicit function of $\vec{A}$ and $\vec{B}$. Interchanging dot and cross in the scalar triple product $(\partial\mathcal{L}_{\rm nc}/\partial\vec{B})\dotv\curl\delta\vec{A}$ and integrating by parts we find
\begin{equation}\label{eq:LncAfc}
	\frac{\delta L_{\rm nc}}{\delta\vec{A}} = \frac{\partial\mathcal{L}_{\rm nc}}{\partial\vec{A}} 
	+ \curl\frac{\partial\mathcal{L}_{\rm nc}}{\partial\vec{B}} \;.
\end{equation}
The corresponding Euler--Lagrange equation is found by setting $\delta L_{\rm nc}/\delta\vec{A}$ to zero.

\subsection{Example: Ideal MHD PSL with cross helicity constraint}\label{sec:IMHDPSL}

In this section we test the PSL approach against the same problem for which the CSL gave unphysical results in Sec.~\ref{sec:IMHDCSL}. Thus we take as Hamiltonian \eqref{eq:Hstd} with the potential energy density term $\mathcal{V} = p/(\gamma - 1) + \vec{B}\dotv\vec{B}/2\muSI - \nu\vec{u}\dotv\vec{B}/\muSI$, the cross-helicity constraint term $\nu K_\Omega^{\rm X}[\vec{u},\vec{B}]$ being subtracted from the \emph{Hamiltonian} rather than adding $\nu K_\Omega^{\rm X}[\vec{v},\vec{B}]$ to the CSL \emph{Lagrangian}. (Constraining the Hamiltonian is more relevant to the constrained energy minimization idea behind the present paper than constraining the CSL action.)  The Lagrangian velocity $\vec{v}$ remains holonomically constrained as in Sec.~\ref{sec:IMHDCSL}, as do $\rho$, $p$, and $\vec{B}$, so $\bm{\uplambda}$ and the structure matrices in \eqref{eq:strucmats1}  remain unchanged. 

Then the PSL density is [cf. \eqref{eq:MHDCSL}]
\begin{align}
	\mathcal{L}_{\rm nc} 
	&= \uplambda_1\vec{u}\dotv\vec{v} - \frac{\uplambda_1\vec{u}^2}{2} \nonumber\\
	&\quad - \frac{\uplambda_2}{\gamma - 1} - \frac{\bm{\uplambda}_3\dotv\bm{\uplambda}_3}{2\muSI} 
	+ \nu\frac{\vec{u}\dotv\bm{\uplambda}_3}{\muSI} \;, \label{eq:MHDPSL}
\end{align}
which now has $\vec{u}$ as a free field, so $\bm{\upeta} = [\vec{u}]$. 

\begin{figure}[htbp] 
\centering
\includegraphics[width=8.5cm]{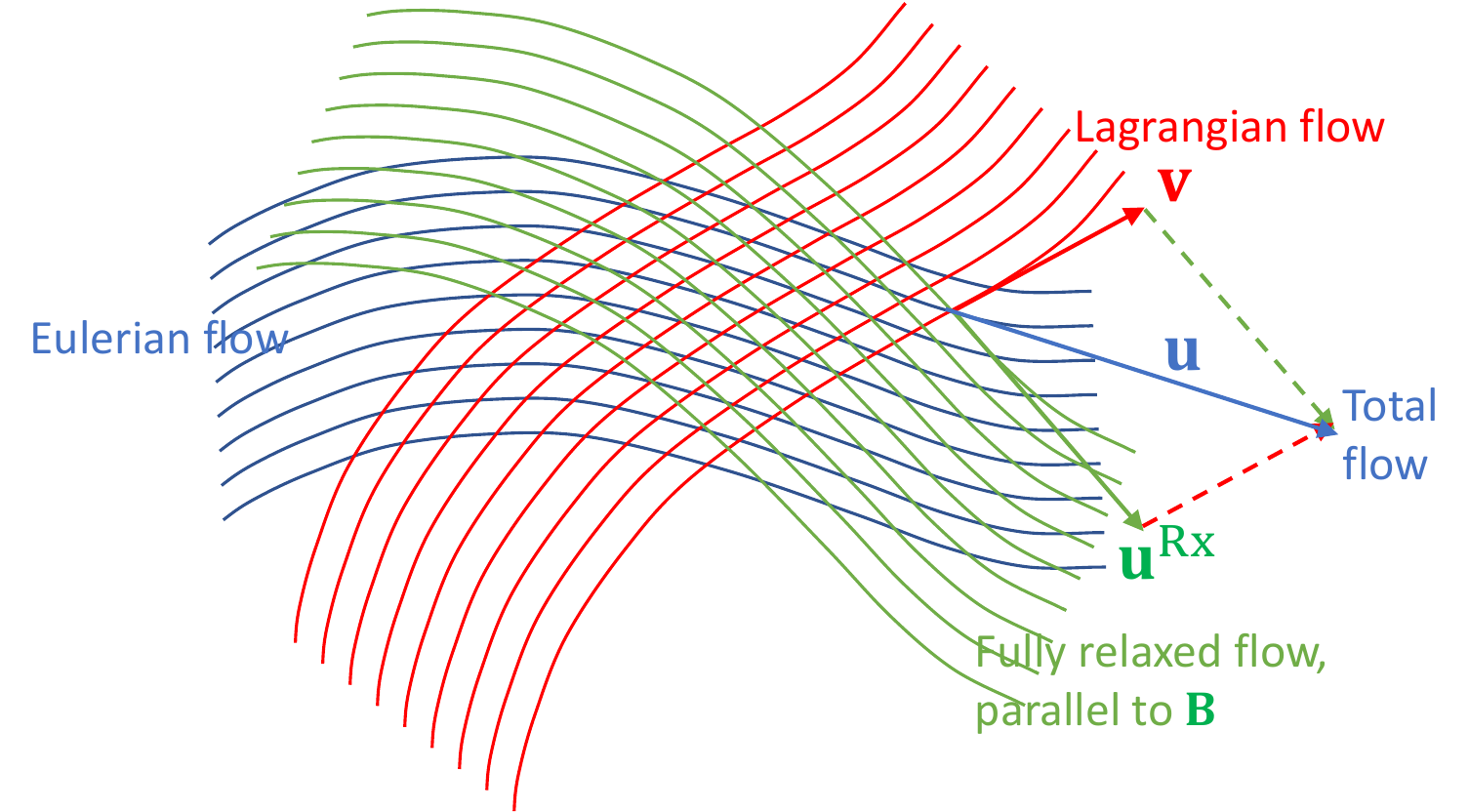}
\caption{Cartoon of the relationship between the three flow velocities appearing in the Euler--Lagrange equation \eqref{eq:vXshift}: the relaxed, field-aligned background flow $\vec{u}^{\rm Rx}$ (green), the perturbing, relabeling  flow $\vec{v}$ (red), and the resultant Eulerian flow $\vec{u}$ (blue). (Color online.)}
\label{fig:ThreeFlows}
\end{figure} 

However, there are no $\bm{\upeta}_t$ or $\grad\bm{\upeta}$ terms so the Euler--Lagrange equation \eqref{eq:etaEL} becomes simply $\partial\mathcal{L}_{\rm nc}/\partial\vec{u} = \bm{0}$, giving (after dividing by $\lambda_1$)
\begin{equation}\label{eq:vXshift}
	\vec{u} = \vec{v} + \frac{\nu\bm{\uplambda}_3}{\lambda_1\muSI} 
	\equiv \vec{v} + \vec{u}^{\rm Rx} \;,
\end{equation}
where the magnetic-field-aligned velocity
\begin{equation}\label{eq:ufullRx}
	\vec{u}^{\rm Rx}\equiv \frac{\nu \vec{B}}{\muSI\rho}
\end{equation}
is the \emph{fully relaxed} flow velocity, found in Appendix~\ref{sec:genTaylor} to result from the cross helicity constraint when extremizing the Hamiltonian $H_\Omega^{\rm Rx}$, \eqref{eq:HRx}. Figure~\ref{fig:ThreeFlows} gives a visualization of \eqref {eq:vXshift}, showing the flow $\vec{u}$ (blue) as the vector sum of the flow $\vec{v}$ (red) and the field-aligned background flow $\vec{u}^{\rm Rx}$  (green), (Color online.)

When $\nu = 0$, $\vec{u}^{\rm Rx} = 0$ also, and we may then identify $\vec{u}$ and $\vec{v}$. However, adding the cross-helicity constraint makes the velocity-like noncanonical momentum field $\vec{u}$ and the Lagrangian-map-constrained velocity $\vec{v}$ \emph{different}. Which is the ``true'' physical fluid velocity?

As a first step toward answering this question, we consider the fluid equation of motion in momentum conservation form---\eqref{eq:Gdef} gives
\begin{align}
	\vec{G} & = \uplambda_1\vec{u}
	\equiv \rho\vec{u} 
	\;,
	\label{eq:GMHDPSL} 
\end{align}
and \eqref{eq:Tdef} gives
\begin{align}
	\vsf{T} &= 
	\lambda_1\vec{v}\vec{u}
	+ \left[0,0,\bm{\uplambda}_3\right] \left[\frac{v^2}{2},\, \frac{1}{\gamma - 1},\, 
	\frac{\nu\vec{u} - \bm{\uplambda}_3}{\muSI}\right]^\trnsp \nonumber\\
	&\quad + \vsf{I}\left(\uplambda_1\vec{u}\dotv\vec{v} - \frac{\uplambda_1\vec{u}^2}{2} - \frac{\uplambda_2}{\gamma - 1} 
		- \frac{\bm{\uplambda_3}\dotv\bm{\uplambda_3}}{2\muSI} + \nu\frac{\vec{u}\dotv\bm{\uplambda_3}}{\muSI} \nonumber\right.\\
	&\qquad\left.\mbox{}-\uplambda_1\vec{u}\dotv\vec{v}  + \frac{\uplambda_1\vec{u}^2}{2} 
	+ \gamma\frac{\uplambda_2}{\gamma - 1} 
	      - \frac{\bm{\uplambda}_3\dotv(\nu\vec{u} - \bm{\uplambda}_3)}{\muSI} \right)
	\nonumber\\ %
	&\equiv \rho\vec{v}\vec{u} + \frac{\nu\vec{B}}{\muSI}\vec{u}
	+ \vsf{I}\left(p + \frac{\vec{B}\dotv\vec{B}}{2\muSI}\right) - \frac{\vec{B}\vec{B}}{\muSI} \nonumber\\
	&= \rho\,\vec{u}\vec{u} 
	+ \vsf{I}\left(p + \frac{\vec{B}\dotv\vec{B}}{2\muSI}\right) - \frac{\vec{B}\vec{B}}{\muSI}
	\;. \label{eq:TMHDPSL}
\end{align}

\noindent \textbf{Demonstration that} $\vec{u}$ \textbf{is the IMHD flow velocity:}
\begin{itemize}
\item Unlike the CSL approach in Sec.~\ref{sec:IMHDCSL} the PSL method gives the physically correct momentum density and stress tensor, Eqs.~(\ref{eq:GMHDPSL}) and (\ref{eq:TMHDPSL}), even with the redundant cross-helicity constraint, \emph{provided we identify $\vec{u}$, not $\vec{v}$, as the physical flow velocity}.
\item From \eqref{eq:vXshift}, $\divv(\rho\vec{v}) = \divv(\rho\vec{u})$, so, from \eqref{eq:Continuity}, $\vec{u}$ obeys the required mass 
continuity equation $\partial_t\rho\, + \divv(\rho\vec{u}) = 0$. 
\item Provided $p$ is barotropic (i.e. $p/\rho^\gamma = \const$ throughout $\Omega$), then the required adiabatic pressure equation of motion, \eqref{eq:AdiabaticPressure}, must be satisfied even when $\vec{v}$ is replaced by $\vec{u}$.
\item From \eqref{eq:vXshift}, $\vec{v}\cross\vec{B} = \vec{u}\cross\vec{B}$, so, from \eqref{eq:FrozenFlux}, the required ``frozen-in flux'' equation, $\partial_t \vec{B} = \curl(\vec{u}\cross\vec{B})$, is satisfied. $\quad\Box$
\end{itemize} 
\begin{figure}[htbp] 
\centering
\includegraphics[width=6.5cm]{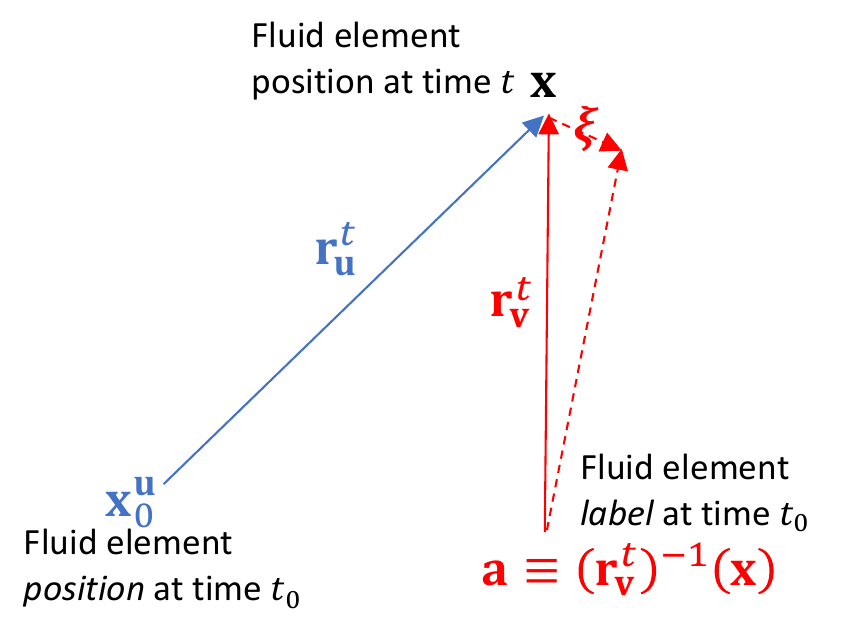}
\caption{Sketch of how the inverse Lagrangian $\vec{v}$-flow map (red) generates labels, $\vec{a}$, for fluid elements in the $\vec{u}$-flow (blue). When making variations about points $\vec{x}$ through displacements $\bm{\xi}$, $\vec{a}$ is held fixed; but it is advected by the $\vec{u}^{\rm Rx}$ flow---see text. (Color online.)}
\label{fig:vLabel}
\end{figure} 

\noindent \textbf{Demonstration that} $\vec{u}^{\rm Rx}$ \textbf{is a reference flow:}

As depicted in Fig.~\ref{fig:vLabel}, we can now interpret the Lagrangian velocity $\vec{v}$ as an auxiliary dynamical vector field whose inverse flow to $t = t_0$ gives a labeling, $\vec{a} \equiv (\vec{r}_{\vec{v}}^t)^{-1}$, with the $\vec{v}$-flow map $\vec{r}_{\vec{v}}^t$ being defined in \eqref{eq:vFlow}: \emph{Adding the cross-helicity constraint generates a relabeling transformation.} As all the fluid elements must be advected by the $\vec{u}$-flow, i.e. $\vec{\dot{x}} = \vec{u}$, the labels $\vec{a}$ must clearly \emph{move} to track these fluid elements. \footnote{In that there was no appeal to an initial state, the concept of moving labels was also used, implicitly, in Ref.~\onlinecite{Dewar_70} in the context of waves on a Lagrangian-mean background flow.}

What is the physical interpretation of $\vec{a}$? By definition, $\vec{x} = \vec{r}_{\vec{v}}^t \comp\vec{a}(\vec{x},t)$, so, taking the total time derivative of both sides along the path of a fluid element,
\begin{equation*}
	\vec{u} = \vec{v} + \vec{\dot{a}}\dotv\grad_{\vec{a}}\vec{r}_{\vec{v}}^t(\vec{a}) \;.
\end{equation*}
From \eqref{eq:ufullRx}, $\vec{u} - \vec{v} = \vec{u}^{\rm Rx}$. Thus, dotting from the right with the inverse of the dyadic $\grad_{\vec{a}}\vec{r}_{\vec{v}}^t$,
\begin{equation}\label{eq:identv}
	\vec{\dot{a}} = \vec{u}^{\rm Rx}\dotv[\grad_{\vec{a}}\vec{r}_{\vec{v}}^t(\vec{a})]^{-1} \;.
\end{equation}

Recall the well-known relations (e.g. Ref.~\onlinecite[Eqs. (2.20) and (2.22)]{Newcomb_62} $\vec{B} = \vec{B}_0\dotv\grad_{\vec{a}}\vec{r}_{\vec{v}}^t /\mathcal{J}$ and $\rho = \rho_0/\mathcal{J}$, where subscript $0$ means ``at time $t_0$ and initial position $\vec{a}$,'' and $\mathcal{J}$ is the Jacobian $\det(\grad_{\vec{a}}\vec{r}_{\vec{v}}^t)$. In \eqref{eq:ufullRx} these give
\begin{equation}\label{eq:ufullRx0rel}
	\vec{u}^{\rm Rx} = \vec{u}^{\rm Rx}_0\dotv\grad_{\vec{a}}\vec{r}_{\vec{v}}^t(\vec{a}) \;,
\end{equation}
which integrates the equation of motion for $\vec{u}^{\rm Rx}$: From its definition, \eqref{eq:ufullRx}, and assuming $\partial\Omega$ and hence $\nu$ constant in time, we have, using the evolution equations for $\rho$ and $\vec{B}$, Eqs. (\ref{eq:Continuity}) and (\ref{eq:FrozenFlux}), respectively,
\begin{align}
	\partial_t\vec{u}^{\rm Rx} &= \frac{\nu}{\muSI}\left(\frac{\partial_t\vec{B}}{\rho} - \frac{\vec{B}\partial_t\rho}{\rho^2}\right) \nonumber\\
	&= \frac{\nu}{\muSI}\frac{ -\vec{B}\divv\vec{v} + \vec{B}\dotv\grad\vec{v} - \vec{v}\dotv\grad\vec{B} 
	+ \vec{B}(\divv\vec{v} + \vec{v}\dotv\grad\ln\rho)}{\rho} \nonumber\\
	&= \frac{\nu}{\muSI}\frac{ \vec{B}\dotv\grad\vec{v} - \vec{v}\dotv\grad\vec{B} 
	+ \vec{v}\dotv\grad\ln\rho}{\rho} \nonumber\\
	&= \vec{u}^{\rm Rx}\dotv\grad\vec{v} - \vec{v}\dotv\grad\vec{u}^{\rm Rx} \;. \label{eq:ptltderivurx}
\end{align}

Substituting \eqref{eq:ufullRx0rel} in \eqref{eq:identv} we now have a full identification $\vec{\dot{a}} = \vec{u}^{\rm Rx}_0$, or, more explicitly,
\begin{equation}\label{eq:identv0}
	(\partial_t + \vec{u}\dotv\grad)\vec{a}(\vec{x},t) = \vec{u}^{\rm Rx}(\vec{a},t_0) \;.
\end{equation}
This identifies $\vec{u}^{\rm Rx}$ as the background or reference flow with respect to which the relative velocity $\vec{v}$ is defined. As $t_0$ is arbitrary, we can, at any time $t$, choose $t_0 = t$. In this case $\vec{a} = \vec{x}$ and $\vec{\dot{a}}(\vec{x},t) = \vec{u}^{\rm Rx}(\vec{x},t)$.$ \:\Box$

\section{Elevation of relaxation to a dynamical theory}\label{sec:vRxph}

\subsection{PSL for RxMHD}\label{sec:RxMHDPSL}

In Appendix~\ref{sec:genTaylor} we verify that constrained noncanonical Hamiltonian $H_\Omega^{\rm Rx}$, \eqref{eq:HRx}, is an energy functional whose stationary points under variation give a subset of the solutions of isothermal, ideal magnetohydro\emph{statics} (IMHS), i.e. they meet the ideal Consistency Principle. Thus they are acceptable relaxed solutions. 

However, there are two motivations to generalize the result of Appendix~\ref{sec:genTaylor} to a relaxed magnetohydro\emph{dynamics}: (a) the flow $\vec{u}$ in this ``fully relaxed equilibrium'' is limited to the magnetic-field-aligned flow $\vec{u}^{\rm Rx}$, which is overly restrictive for some purposes even in equilibrium studies; and (b) a time-dependent theory could address a wider class of physical phenomena, such as waves and instabilities.

As $H_\Omega^{\rm Rx}$ gives a satisfactory relaxed magnetostatics it is a natural starting point for a dynamical theory. To do this we replace the Hamiltonian in \eqref{eq:PSLnc} with $H_\Omega^{\rm Rx}[\vec{u}]$ to form the relaxed PSL
\begin{equation}\label{eq:Lphx}
\begin{split}
	L_\Omega^{\rm Rx}[\vec{u},\vec{v}] = \!\int_\Omega\!\! \rho\vec{u}\dotv\vec{v}\,\d V  -  H_\Omega^{\rm Rx} \, ,
\end{split}
\end{equation}
where $H_\Omega^{\rm Rx}$ is as given in \eqref{eq:HRx}. Then the PSL density is [cf. \eqref{eq:MHDCSL}]
\begin{align}
	\mathcal{L}_\Omega^{\rm Rx} 
	&= \rho\vec{u}\dotv\vec{v} - \frac{\rho\vec{u}^2}{2} - \frac{p}{\gamma - 1} - \frac{\vec{B}\dotv\vec{B}}{2\muSI} 	\nonumber\\
	& \quad + \tau_\Omega \frac{\rho}{\gamma - 1} \ln\left(\kappa\frac{p}{\rho^\gamma}\right)
	 + \mu_\Omega\frac{\vec{A}\dotv\vec{B}}{2\muSI} + \XHelmult\frac{\vec{u}\dotv\vec{B}}{\muSI}\;, \label{eq:RxPSL}
\end{align}

In the notation of Sec.~\ref{sec:HolonomicFreeSplit}, relaxation is implemented by moving all but $\rho$ to the set of free variables, i.e. the holonomic variables array is just $\bm{\uplambda} = [\rho]$, and the free variables array is $\bm{\upeta}^\trnsp = [p,\vec{u},\vec{A}]$. The structure matrices become trivial, $\vsf{V} = [0]$, $\vsf{\Lambda} = [1]$.

\subsection{RxMHD Euler--Lagrange equations}\label{sec:RxMHDEL}
The $\vec{u}$ component of \eqref{eq:etaEL} gives, as in Sec.~\ref{sec:IMHDPSL},
\begin{equation}\label{eq:uHamPrEL}
	\rho\vec{u} - \rho\vec{v} = \XHelmult \frac{\vec{B}}{\muSI} \equiv \rho\vec{u}_\Omega^{\rm Rx} \;, 
\end{equation}
where $\vec{u}^{\rm Rx}$ is defined in \eqref{eq:ufullRx}, with $\nu$ set to $\nu_\Omega$ (also see Figure~\ref{fig:ThreeFlows}). Note that $\divv(\rho\vec{u}_\Omega^{\rm Rx}) = 0$.

The $\vec{B}$ component gives, using \eqref{eq:LncAfc}, the \emph{modified Beltrami equation},
\begin{equation}\label{eq:AHamPrEL}
	\curl\vec{B} = \mu_\Omega  \vec{B} + \XHelmult\curl\vec{u}  \;,
\end{equation}
and the final, $p$, component of \eqref{eq:etaEL} gives the \emph{isothermal equation of state},
\begin{equation}\label{eq:pHamPrEL}
	p = \tau_\Omega  \rho  \;.
\end{equation}

From \eqref{eq:deltaLbydeltar},
\begin{align}
	\frac{\delta L_\Omega^{\rm Rx}}{\delta\vec{r}} &= 
	\divv\!\left(\vsf{I}\, \rho\frac{\partial\mathcal{L}_\Omega^{\rm Rx}}{\partial\rho}
		\right)
	- \frac{\partial\mathcal{L}_\Omega^{\rm Rx}}{\partial\rho}\grad\rho \nonumber \\
	&= \rho \grad\left(\vec{u}\dotv\vec{v} - \frac{\vec{u}^2}{2}  - \tau_\Omega\ln\frac{\rho}{\rho_\Omega}\right)
	\label{eq:deltaLRxbydeltar}\;,
\end{align}
where $\rho_\Omega$ is an arbitrary spatial constant.
The momentum equation is, from \eqref{eq:rEL},
\begin{align}
&	\partial_t(\rho\vec{u}) 
		+\divv(\rho\vec{v}\vec{u})
	 	+(\grad\vec{v})\dotv\rho\vec{u} = \frac{\delta L_\Omega^{\rm Rx}}{\delta\vec{r}} \nonumber \\
&\qquad = \rho \grad\left(\vec{u}\dotv\vec{v} - \frac{\vec{u}^2}{2}  - \tau_\Omega\ln\frac{\rho}{\rho_\Omega}\right) \;,\quad\text{hence}
	\; \nonumber \\
& \partial_t(\rho\vec{u}) + \divv(\rho\vec{v}\vec{u})  - \rho(\grad\vec{u})\dotv\vec{v} =
	-\rho\grad h_\Omega \;, \label{eq:xiHamPrEL}
\end{align}
where $h_\Omega$ is as defined in \eqref{eq:Head}, $u^2/2 + \tau_\Omega\ln\rho/\rho_\Omega\;$.

Taking the divergence of both sides of \eqref{eq:uHamPrEL} we have $\divv(\rho\vec{v}) = \divv(\rho\vec{u})$, so $\vec{u}$ obeys the same continuity equation as $\vec{v}$, \eqref{eq:Continuity}. That is,
\begin{equation}\label{eq:ucont}
	\partial_t\rho + \divv(\rho\vec{u}) = 0 \;.
\end{equation}

We can condense \eqref{eq:AHamPrEL} by writing it in terms of the \emph{vorticity}, $\bm{\omega} \equiv \curl\vec{u}$, giving $\curl\vec{B} = \mu_\Omega\vec{B}  + \XHelmult \bm{\omega}$. Further physical insight is gained by writing \eqref{eq:AHamPrEL} in terms of electric current $\vec{j} \equiv \curl\vec{B}/\muSI$,
\begin{align}
	\vec{j} = \frac{\mu_\Omega}{\muSI} \vec{B} + \frac{\XHelmult}{\muSI} \bm{\omega}
	\label{eq:AEnPrELaltj}\;,
\end{align}
the first term on the RHS of \eqref{eq:AEnPrELaltj} being the usual parallel electric current of the linear-force-free (Beltrami) magnetic field model while the second term is a \emph{vorticity-driven current}. \cite{Yokoi_2013}

The equation of motion \eqref{eq:xiHamPrEL} can be written, using the mass conservation equation, \eqref{eq:Continuity}, and dividing by $\rho$,
\begin{equation}\label{eq:xiHamPrELomega1}
	\partial_t\vec{u}  +\bm{\omega}\cross\vec{v} = - \grad h_\Omega \;.
\end{equation}
Taking the curl of both sides gives $\partial_t\bm{\omega} + \curl(\bm{\omega}\cross\vec{v}) = 0$. Thus, in steady flow there must exist a potential, $ \homvOmega$ say, such that $\bm{\omega}\cross\vec{v} = \grad \homvOmega$, which implies $\grad(h_\Omega + \homvOmega) = 0$. That is, \emph{any} steady RxMHD state has a generalized Bernoulli equation $h_\Omega + \homvOmega = \const$. Note that $\vec{v}\dotv\grad\homvOmega = \bm{\omega}\dotv\grad\homvOmega = 0$, so generically the streamlines of $\vec{v}$ and the vorticity lines of $\vec{u}$ must either lie within invariant tori of these two flows or occupy chaotic regions in which $\homvOmega = \const$.

Eliminating $\vec{v}$ using \eqref{eq:AEnPrELaltj}, this equation can also be written as
\begin{equation}\label{eq:xiHamPrELu}
	\rho(\partial_t + \vec{u}\dotv\grad)\vec{u} = - \grad p + \vec{j}\cross\vec{B} \;.
\end{equation}
Thus \eqref{eq:xiHamPrELu} is simply the standard IMHD equation of motion, \eqref{eq:idealEqMot}, implying that our \emph{RxMHD equation of motion is consistent with Newton's second law}. 

Note that \eqref{eq:xiHamPrELu}, and hence \eqref{eq:xiHamPrEL}, can also be written in the standard conservation form [see \eqref{eq:GMHDPSL}, \eqref{eq:TMHDPSL}, and \eqref{eq:momcon}]
\begin{equation}\label{eq:xiHamPrELuconservation}
	\partial_t(\rho\vec{u}) + \divv\left[\rho\vec{u}\vec{u} + \left(p + \frac{B^2}{2\muSI}\right)\Ident - \frac{\vec{B}\vec{B}}{\muSI}\right] = 0 \;.
\end{equation}

\subsection{On the RxMHD equations of motion}\label{sec:RxMHDuB}

In the limit $\XHelmult \to 0$, \eqref{eq:uHamPrEL} shows that $\vec{v} = \vec{u}$ and the Euler--Lagrange equations become exactly the same as those in the original dynamical MRxMHD paper \cite{Dewar_Yoshida_Bhattacharjee_Hudson_2015} --- they describe uncoupled Beltrami magnetic fields and Euler flows. In this case, the phase-space Lagrangian method is equivalent to the configuration-space Lagrangian method used in Ref.~\onlinecite{Dewar_Yoshida_Bhattacharjee_Hudson_2015}. The physical implication is that, in this limit, both methods describe relaxation of magnetic field, but not fluid (unless we set $\vec{v} = 0$, in which case \eqref{eq:uHamPrEL} gives the same magnetic-field-aligned $\vec{u}$ as the relaxed equilibrium flow given in Appendix~ \ref{sec:genTaylor}). 

To understand the mathematical nature of RxMHD when $\XHelmult \neq 0$ within a given domain $\Omega$, with boundary $\partial\Omega$, we distinguish between the \emph{dynamical variables} $\vec{u}$ and $\rho$, whose time evolution is to be found by solving equations of motion, and \emph{passive variables}, whose time evolution depends only on the time dependence of $\partial\Omega$, like the Lagrange multipliers $\tau_\Omega,\mu_\Omega,\text{and }\XHelmult$, or fields whose time dependence is, in addition, driven implicitly by that of the dynamical variables. The main example of the latter class is the magnetic field, a functional of $\vec{u}$ found by solving the inhomogeous modified Beltrami equation, \eqref{eq:AHamPrEL}, to give  
\begin{equation}\label{eq:AEnPrELinv}
	\vec{B} =  \vec{B}_{\Psi} + \XHelmult (\mathbf{curl} - \mu_\Omega\Ident)^{-1}\dotv\, \curl\vec{u}\;,
\end{equation}
where $\vec{B}_{\Psi}$ is the unique solution of the \emph{homogeneous} Beltrami equation, \eqref{eq:AHamPrEL} with $\XHelmult = 0$, given prescribed magnetic fluxes $\Psi$, a set of constants of the motion whose number depends on the topological genus of $\Omega$. \cite{Yoshida_Giga_90,Yoshida_Dewar_12} (In the above we assumed $\mu_\Omega $ is not an eigenvalue of the homogeneous Beltrami equation.)

Likewise, the pressure $p$ is known in terms of $\rho$ through \eqref{eq:pHamPrEL}, so \eqref{eq:ucont} and \eqref{eq:xiHamPrELu} constitute an infiniite dimensional dynamical system of the form $\partial_t(\rho,\vec{u}) = \textsf{\textbf{f}}[\rho,\vec{u}]$.

The field $\vec{v}$ allows freedom for the initial conditions for $\vec{u}$ to be specified arbitrarily, through \eqref{eq:uHamPrEL}, rather than to be constrained to the fully relaxed, field-aligned flow $\XHelmult\vec{B}/\muSI\rho$, but it should not be regarded as giving cross-field flow only. For example, \eqref{eq:vFA} shows $\vec{v}$ with both cross-field and field-aligned flow.

We could in principle display the dynamics in terms of $\vec{v}$, instead of $\vec{u}$, but it is considerably more complicated and difficult to interpret. However, a hybrid approach, where one first specifies $\vec{v}$ and then seeks compatible solutions for $\rho$, $\vec{u}$, and $\vec{B}$, can restrict attention to interesting classes of solutions.

For instance, if we take \vec{v} to be \emph{purely} field-aligned in such as way as to counteract the fully relaxed flow, i.e. by setting $\vec{v} = -\XHelmult\vec{B}/\muSI\rho$, then \eqref{eq:uHamPrEL} shows $\vec{u} = 0$ and we recover the Taylor-relaxed state as a special case. 

More interesting are\\ 
\noindent\textbf{Solutions with a continuous symmetry}:\\
Suppose the boundary of $\Omega$ possesses a continuous geometric symmetry and seek solutions, equilibrium or possibly dynamical, that maintain this symmetry in time.

For specificity, consider the important case of axisymmetric systems, in which scalar quantities are independent of the toroidal angle $\phi$, so that, for example, $\esub{\phi}\dotv\grad\rho = 0$, where $\esub{\phi}(\phi)$is the unit vector $R\grad\phi$, $R$ being the distance from the symmetry, $Z$, axis. 

Similarly, $\esub{\phi}\dotv\grad h_\Omega = 0$, so dotting both sides of \eqref{eq:xiHamPrELomega1} with $\esub{\phi}$ gives
\begin{equation}\label{eq:xiHamPrELomega1v}
	\partial_t(\esub{\phi}\dotv\bm{\omega}) + \esub{\phi}\dotv\,\bm{\omega}\cross\vec{v} = 0 \;.
\end{equation}
If we now choose 
\begin{equation}
 	\vec{v} = |\vec{v}|\esub{\phi} 
\end{equation}
then \eqref{eq:xiHamPrELomega1v} is satisfied for any $\vec{u}$ solution such that the toroidal component of vorticity, $\esub{\phi}\dotv\bm{\omega}(\vec{x})$, is constant in time throughout $\Omega$. An example of of such a solution is examined in detail in Appendix~\ref{sec:RxMHDsteady}.

\section{Linearized dynamics in the WKB approximation}\label{sec:WKBRxMHD}
\subsection{Linearization}\label{sec:LinRxMHD}

As a first step toward understanding the dynamical implications of the RxMHD equations, we linearize around a steady flow ($\partial_t \mapsto 0$) solution of  Eqs.~(\ref{eq:uHamPrEL}), (\ref{eq:ucont}), (\ref{eq:AEnPrELaltj}), and (\ref{eq:xiHamPrELomega1}), in a domain $\Omega$ with either fixed boundaries or with only low-amplitude, short-wavelength perturbations. Thus, insert in these equations the ansatz $\vec{u} = \vec{u}^{(0)} + \alpha\vec{u}^{(1)} +O(\alpha^2)$, and $\vec{v} = \vec{v}^{(0)} + \alpha\vec{v}^{(1)} + O(\alpha^2)$, and similarly for $\rho^{(0)}$, where $\alpha$ is the amplitude expansion parameter (for an example of an equilibrium with nonzero $\vec{u}^{(0)}$ and $\vec{v}^{(0)}$ see Appendix~\ref{sec:RxMHDsteady}). For fixed boundaries, or for short-wavelength perturbations, the entropy, helicity and cross-helicity integrals are conserved at $O(\alpha)$, with therefore no perturbation in the Lagrange multipliers. Thus here we take $\tau_\Omega$, $\mu_\Omega$, and $\XHelmult$ as time-independent constants. Also, from here on we take the superscript (0) to be implicit, e.g. $\rho$ means $\rho^{(0)}$ etc. 

For short wavelength, high frequency velocity perturbations we use the eikonal ansatz
\begin{equation}\label{eq:eikonvRxHD}
	\vec{u}^{(1)} = \widetilde{\vec{u}}(\vec{x},t)\exp\left(\frac{\ij \varphi(\vec{x},t)}{\varepsilon}\right) \;,
\end{equation}
with similar notations for linear perturbations of other quantities, $\varepsilon$ being the WKB (local plane-wave) expansion parameter. The instantaneous local values of frequency and wave vector are then defined as $\omega(\vec{x},t) \equiv -\partial_t\varphi$ and $\vec{k} \equiv \grad\varphi$. Taking $\varphi$ and equilibrium quantities to vary on $O(1)$ spatial and temporal scales, $\omega$, $\vec{k}$, $\partial_t\vec{u}$, $\grad\vec{u}$, $\mu_\Omega$, $\XHelmult$ etc. are $O(1)$, but $\partial_t\vec{u}^{(1)}$, $\grad\vec{u}^{(1)}$ etc. are large, $O(\varepsilon^{-1})$.

\subsection{Short-wavelength IMHD perturbations}\label{sec:WKBIMHD}
Dynamical relaxation theory is physically applicable on slow, quasi-equilibrium timescales $t \gtrsim \tau_{\rm Rx}$. In the opposite limit of fast dynamics, on times $t \ll \tau_{\rm Rx}$, of perturbations on relaxed equilibria (which are a subset of ideal equilibria) or slowly evolving states, it is more physically consistent to use IMHD than RxMHD. 

The three IMHD plane-wave branches are derived in many text books. We shall follow Ref.~\onlinecite[pp. 172--173]{Hosking_Dewar_2015}, where the local eigenvalue equation for IMHD waves is given in the form
\begin{equation} \label{eq:magnetic4}
    {\bm{\mathsf D}} \dotv \widetilde{\vec{u}} = 0 \; ,
\end{equation}
where, in the isothermal ($\gamma = 1$) case,
\begin{equation} \label{eq:D}
\begin{split}
    {\bm{\mathsf D}} &\equiv  \rho\, \omega'^{2}\;
    {\bm{\mathsf I}}- p \,{\bf k} {\bf k} \\
    & \quad - \muSI^{-1} ({\bf k B} - {\bf k} \dotv {\bf B}\; {\bm{\mathsf I}}) \dotv ({\bf B k}
    - {\bf k} \dotv {\bf B}\; {\bm{\mathsf I}})\;.
\end{split}
\end{equation}
with $\omega'$ denoting the Doppler-shifted frequency in a local frame moving with the fluid, i.e. $\omega' \equiv \omega - \vec{k}\dotv\vec{u}$. The local dispersion relation may be found by representing $\vsf{D}$ as a matrix using the co and contravariant bases $\{\esub{i}\}$, $\{\esup{i}\}$,
\begin{equation} \label{eq:kBbasis}
\begin{split}
    \esub{1} &\equiv \vec{k} = k^2\esup{1} + \vec{k}\dotv\vec{B}\, \esup{2} \\
    \esub{2} & \equiv \vec{B}  = \vec{k}\dotv\vec{B}\,\esup{1} + B^2\,\esup{2} \\
    \esub{3} & \equiv |\vec{k}\cross\vec{B}|^2\, \esup{3}
\end{split}
\end{equation}
and setting $\det\vsf{D} = 0$. The resulting dispersion relation has the three roots
\begin{subequations}
    \begin{alignat}{2}
    &\text{Alfv\'en waves:}&
    \omega'^{2} &= k_\parallel^2 c^{2}_{\rm A} \;,\label{eq:Alfvendisprel}\\
    &\text{slow MS waves:}\: &
    \omega'^{2} &= \frac{1}{2} k^{2}
    \left( c^{2}_{\rm s} + c^{2}_{\rm A} \right) (1-\sqrt{1-\alpha^2}\,) \nonumber\\
    &&& \approx k_\parallel^2 c_{\rm s}^2 \;,\label{eq:SMSdisprel} \\
    &\text{fast MS waves:}&
    \omega'^{2} &= \frac{1}{2} k^{2} \left( c^{2}_{\rm s} +
    c^{2}_{\rm A} \right)(1\, + \,\sqrt{1 - \alpha^2}\,) \nonumber\\
    &&& \approx k^2 c_{\rm A}^2   \;,\label{eq:FMSdisprel} 
    \end{alignat} 
\end{subequations}
where ``MS'' stands for ``magnetosonic,''
\begin{equation}\label{eq:MHDdispalphadef}
	\alpha^2 \equiv 4\,{\frac{k_{\|}^{2}}{k^2}}\; {\frac{c^2_{s} c^{2}_{\rm A}} {(c^{2}_{\rm s}+c^2_A)^{2}}}\;,
\end{equation}
$k_\parallel \equiv \vec{k}\dotv\vec{B}/B$,
$ c_{\rm s} \equiv (p/\rho)^{1/2} = \sqrt{\tau_\Omega}$ is the isothermal sound speed, and
$c_{\rm A} \equiv (B^{2}/\muSI \rho)^{1/2}$
defines the local Alfv\'en speed. The notation $\approx$ refers to the low-$\beta$ approximation $c_{\rm s}/c_{\rm A} \ll 1$. Similar simplifications occur for ``flute-like'' perturbations, ie. when $k_\parallel/k \ll 1.$

\subsection{Short-wavelength RxMHD perturbations}\label{sec:WKBMvRxHD}

The linearizations of Eqs.~(\ref{eq:uHamPrEL}), (\ref{eq:ucont}), (\ref{eq:AEnPrELaltj}), and (\ref{eq:xiHamPrELomega1}) are
\begin{align}
	& \rho\vec{v}^{(1)} + \rho^{(1)}\vec{v} 
	= \rho\vec{u}^{(1)} + \rho^{(1)}\vec{u} - \XHelmult \frac{\vec{B}^{(1)}}{\muSI} \label{eq:uHamPrELlin} \\
	&\partial_t\rho^{(1)}  + \divv(\rho\vec{u}^{(1)} + \rho^{(1)}\vec{u}) = 0 \label{eq:ucontlin} \\
	&\curl\vec{B}^{(1)}  = \mu_\Omega\vec{B}^{(1)} + \XHelmult \bm{\omega}^{(1)}  \label{eq:AEnPrELaltomegalin} \\
	&\partial_t\vec{u}^{(1)}  + \bm{\omega}\cross\vec{v}^{(1)} + \bm{\omega}^{(1)}\cross\vec{v}= - \grad h_\Omega^{(1)} \;, \label{eq:xiHamPrELomega1lin} \\
\text{where  } & h_\Omega^{(1)} = \vec{u}\dotv\vec{u}^{(1)} + \tau_\Omega \frac{\rho^{(1)}}{\rho} \label{eq:Headlin} \;.
\end{align}

Using \eqref{eq:eikonvRxHD} in  Eqs.~(\ref{eq:uHamPrELlin}), (\ref{eq:ucontlin}), (\ref{eq:AEnPrELaltomegalin}), and (\ref{eq:xiHamPrELomega1lin}) gives, to leading order in $\varepsilon$ with the orderings $\mu_\Omega$ and $\XHelmult = O(\varepsilon^0)$,
\begin{align}
	& \widetilde{\vec{v}} 
	=\widetilde{\vec{u}} + (\vec{u} - \vec{v})\frac{\widetilde{\rho}}{\rho} - \XHelmult \frac{\widetilde{\vec{B}}}{\muSI\rho} \nonumber\\ &\:\: 
	=\widetilde{\vec{u}} + \frac{\XHelmult}{\muSI\rho} \left(\frac{\widetilde{\rho}}{\rho}\,\vec{B} - \widetilde{\vec{B}}\right) \label{eq:uHamPrELtilde} \\
	&\frac{\widetilde{\rho}}{\rho}  = \frac{\vec{k}\dotv\widetilde{\vec{u}}}{\omega'}  \label{eq:uconttilde} \\
	&\vec{k}\cross\widetilde{\vec{B}}  = \XHelmult \vec{k}\cross\widetilde{\vec{u}} \label{eq:AEnPrELaltomegatilde}
	\;,\quad \vec{k}\dotv\widetilde{\vec{B}} = 0 \\
	&\omega\widetilde{\vec{u}} = (\vec{k}\cross\widetilde{\vec{u}})\cross\vec{v}
	+ \vec{k}\left(\vec{u}\dotv\widetilde{\vec{u}} + \tau_\Omega \frac{\widetilde{\rho}}{\rho}\right) \nonumber\\
	&\quad\: = (\vec{k}\cross\widetilde{\vec{u}})\cross\left(\vec{u} -\frac{\XHelmult}{\muSI\rho}\vec{B}\right)
	+ \vec{k}\left(\vec{u}\dotv\widetilde{\vec{u}} + \tau_\Omega \frac{\widetilde{\rho}}{\rho}\right) \;, \nonumber\\&\text{i.e.}\:\: 
	\omega'\widetilde{\vec{u}} = \left[\frac{\XHelmult}{\muSI\rho}(\vec{k}\vec{B} - \vec{k}\dotv\vec{B}\,\Ident)
	\dotv\widetilde{\vec{u}} + \tau_\Omega\frac{\vec{k}\vec{k}}{\omega'}\right]\dotv\widetilde{\vec{u}}
	\label{eq:xiHamPrELomega1tilde} 
\end{align}
where $\mu_\Omega\widetilde{\vec{B}}$ and $\bm{\omega}\cross\widetilde{\vec{v}}$ have been dropped as higher order in $\varepsilon$ than other terms in \eqref{eq:AEnPrELaltomegatilde} and \eqref{eq:xiHamPrELomega1tilde}, respectively and $\omega'$ is as in \eqref{eq:D}.

Gathering all terms in \eqref{eq:xiHamPrELomega1tilde} on the LHS and multiplying by $\rho\omega'$ gives the eigenvalue equation
\begin{equation}\label{eq:Rxeigvaleqn}
	\vsf{D}_{\rm Rx} \dotv \widetilde{\vec{u}} = 0 \;,
\end{equation}
where, using \eqref{eq:kBbasis},
\begin{align} 
    \vsf{D}_{\rm Rx} 
    &\equiv  \rho\omega'^2\,\Ident - p\, \vec{k} \vec{k}  
     - \frac{\XHelmult \omega'}{\muSI} (\vec{k B} - \vec{k}\dotv\vec{B}\, \Ident) \label{eq:DRx}  \;.
\nonumber\\      &= \left(\rho\omega'^2 - k^2 p  - \frac{\XHelmult \omega'}{\muSI}\vec{k}\dotv\vec{B}\right)\esub{1}\esup{1} \nonumber\\
     &\qquad\qquad\qquad\qquad - \left(p\,\vec{k}\dotv\vec{B} + \frac{\XHelmult \omega'}{\muSI}B^2\right) \esub{1}\esup{2} \nonumber\\
     &+\left(\rho\omega'^2 + \frac{\XHelmult \omega'}{\muSI}\vec{k}\dotv\vec{B}\right)\left(\esub{2}\esup{2}+ \esub{3}\esup{3}\right)  
\end{align}

There being only one off-diagonal component when expanded in the basis $\esub{i}\esup{j}$, the determinant is the product of the diagonals,
$$\left(\rho\omega'^2 - \frac{\XHelmult\vec{k}\dotv\vec{B}}{\muSI}\omega' - k^2 p\right)
\left(\rho\omega' + \frac{\XHelmult\vec{k}\dotv\vec{B}}{\muSI}\right)^2\omega'^2 \;.$$
Setting this determinant to zero gives the dispersion relations
\begin{align}\label{eq:disprels}
	\omega_1' &= 0,\:\:\omega_2' = -\frac{\XHelmult\vec{k}\dotv\vec{B}}{\muSI\rho},\:
	(\text{i.e.}\: \omega_2 = \vec{k}\dotv\vec{v}),\:\text{and} \nonumber\\
	\omega_{3\pm}' &= \frac{1}{2}\left\{\frac{\XHelmult\vec{k}\dotv\vec{B}}{\muSI\rho}
	\pm\left[\left(\frac{\XHelmult\vec{k}\dotv\vec{B}}{\muSI\rho}\right)^2 + 4k^2 c_{\rm s}^2\right]^{1/2}\right\}
\end{align}
giving the group velocities
\begin{align}
	\frac{\partial\omega_1}{\partial\vec{k}} &= \vec{u}, \:\: \frac{\partial\omega_2}{\partial\vec{k}} = \vec{v}, \quad\text{and}\nonumber\\
	\frac{\partial\omega_{3\pm}}{\partial\vec{k}} 
	&= \frac{1}{2}\frac{\XHelmult\vec{B}}{\muSI\rho}\left\{1
	\pm\frac{\XHelmult\vec{k}\dotv\vec{B}}{\muSI\rho}\left[\left(\frac{\XHelmult\vec{k}\dotv\vec{B}}{\muSI\rho}\right)^2 + 4k^2 c_{\rm s}^2\right]^{-1/2}\right\} \nonumber\\
	& \qquad \pm 2\vec{k}c_{\rm s}^2
	\left[\left(\frac{\XHelmult\vec{k}\dotv\vec{B}}{\muSI\rho}\right)^2 + 4k^2 c_{\rm s}^2\right]^{-1/2}
\label{eq:relaxedEulergrpvels} \;.
\end{align}

In the limit $\XHelmult = 0$ this formulation of relaxed Euler flow gives entropy waves advected by the flow and simple sound waves, uncoupled to $\vec{B}$, as in Ref.~\onlinecite{Dewar_Yoshida_Bhattacharjee_Hudson_2015} and  Ref.~\onlinecite{Dewar_Tuen_Hole_2017} (as expected, since the cross-helicity constraint has been dropped). Even when $\XHelmult \neq 0$ we have clearly eliminated all the IMHD waves, replacing them with four branches, two being waves advected with $\vec{u}$ and $\vec{B}$ and two being  hybrids of simple sound waves and entropy waves. 

We now demonstrate that the ideal Ohm's Law is \emph{not} necessarily respected by these linear waves. We proceed by finding a case where the solvability condition, $\curl(\vec{u}\cross\vec{B}) = 0$, for the electrostatic potential [see \eqref{eq:OhmPhiLaw} in the next section], is not satisfied. For linear waves, the solvability condition becomes $\vec{k}\cross(\vec{u}\cross\widetilde{\vec{B}} + \widetilde{\vec{u}}\cross\vec{B}) = 0$.

From \eqref{eq:AEnPrELaltomegatilde}, $\widetilde{\vec{B}} = \XHelmult(\Ident - \vec{k k}/k^2)\dotv\widetilde{\vec{u}}\;.$ Consider for example the branch of waves satisfying $\omega_1' = 0$, which, from \eqref{eq:Rxeigvaleqn} and \eqref{eq:DRx}, satisfy $\vec{k}\dotv\widetilde{\vec{u}}$, i.e. have transverse polarization in velocity (and, as with all waves, in magnetic field).  Then $\vec{k}\cross(\vec{u}\cross\widetilde{\vec{B}} + \widetilde{\vec{u}}\cross\vec{B}) = \vec{k}\dotv(\vec{B} - \XHelmult\vec{u})\widetilde{\vec{u}}$. This cannot vanish for all $\vec{k}$ unless $\vec{B} = \XHelmult\vec{u}$, which is not in general true. 

This indicates either that we need to invoke turbulent e.m.f.'s during dynamical RxMHD evolution, or to implement an ideal-Ohm constraint as discussed in Sec.~\ref{sec:IOhm}. 
\section{Ideal Ohm's Law and cross-field flow}\label{sec:IOhm} 

The IMHD-equilibrium Consistency Principle requires that \eqref{eq:FrozenFlux} (with $\partial_t\vec{B} = 0$) should be satisfied in a relaxed equilibrium. Thus we take the unqualified term \emph{relaxed MHD equilibrium} to imply that the ``ideal Ohm's Law,'' $\vec{E} + \vec{u}\cross\vec{B} = 0$, is satisfied with, in general, a nonzero $\vec{E}$. 

Given a particular frame (the \emph{Lab frame}) in which an MHD equilibrium appears as a steady state, so that $\curl\vec{E} = -\partial_t\vec{B} = 0$, one can always choose a gauge in which $\vec{E} = -\grad\Phi$, $\Phi$ being a single-valued electrostatic potential. 

[One might object that, taking $\Omega$ to be a simple torus for simplicity, $\Phi$ should also include a secular ``loop voltage'' term, $-E_{\rm ext}\phi/2\pi$, induced by an external time-dependent poloidal magnetic flux linking $\Omega$ ($\phi$ being the geometric toroidal angle and $E_{\rm ext}$ a constant throughout $\Omega$). Noting that $\vec{u}\cross\vec{B}\,\dotv\,\d\vec{l} \equiv 0$ for any line element $\d\vec{l}$ aligned with $\vec{B}$, we see that the line integral  around any closed field line within $\Omega$ vanishes, $-\oint\!\vec{u}\cross\vec{B}\,\dotv\,\d\vec{l} = 0$, while $\oint\!\vec{E}\,\dotv\,\d\vec{l} = N E_{\rm ext}$, $N$ being the number of toroidal turns before the field line closes on itself. Equating the two shows that $E_{\rm ext} = 0$. As (possibly long) closed field lines almost always exist, this shows that $\Phi$ is indeed  generically single valued. Physically, this is a consequence of the assumption of no gaps in the perfectly conducting interfaces, so that the linking fluxes are always conserved no matter what the genus of $\Omega$.]

Thus, in the Lab frame, the equilibrium ideal Ohm's law is electrostatic,
\begin{equation}\label{eq:OhmPhiLaw}
	\grad\Phi = \vec{u}\cross\vec{B} \;, 
\end{equation}
implying $\vec{u} = \vec{u}_\perp + u_\parallel\vec{B}/B$, i.e. $\vec{u}$ is the vector sum of the $\vec{E}\cross\vec{B}$, or \emph{cross-field} flow $\vec{u}_\perp = -\grad\Phi\cross\vec{B}/B^2$ and a parallel flow, $u_\parallel$, not determined by $\Phi$.  Equation~\ref{eq:OhmPhiLaw} also implies $\vec{B}\dotv\grad\Phi = 0$ and $\vec{u}\dotv\grad\Phi = 0$, i.e. that $\Phi = \const$ on both magnetic field and flow lines.  As a consequence, if $\Phi$ has smoothly nested level surfaces, then both $\vec{u}$ and $\vec{B}$ lie in the local tangent plane at each point on each isopotential surface.

Finn and Antonsen Ref.~\onlinecite[after Eq.~(29)]{Finn_Antonsen_83} conclude from this constancy of $\Phi$ along a field line that ``if the turbulent relaxation has ergodic field lines throughout the plasma volume,'' then $\grad\Phi = 0$, which implies from \eqref{eq:OhmPhiLaw} that $\vec{u}\cross\vec{B} = 0$. We shall call such field-aligned steady flows \emph{fully relaxed equilibria}. In Appendix~\ref{sec:genTaylor} we find that stationary points of the MHD energy, subject only to microscopic mass and macroscopic entropy, magnetic helicity and cross helicity constraints, are indeed fully relaxed equilibria.

However the Consistency Principle applies only to the final, non-turbulent, state of relaxation, where it seems highly unlikely that field lines could ever fill the whole of $\Omega$ ergodically, though in fully three dimensional plasmas with islands this may be a good model for chaotic separatrix subregions. In most cases the class of fully relaxed equilibria seems unnecessarily restrictive. Indeed, it does not include many equilibria of physical interest, in particular, tokamaks with strong toroidal flow. 

Thus Finn and Antonsen Ref.~\onlinecite[Sec. III]{Finn_Antonsen_83} go on to construct an axisymmetric equilibrium \emph{with} cross-field flow by adding the additional constraint of conservation of angular momentum in the relaxed energy principle. As their equilibrium satisfies \eqref{eq:OhmPhiLaw} it satisfies our IMHD Consistency Principle so it definitely qualifies as a relaxed MHD equilibrium. In Appendix~\ref{sec:RxMHDsteady} we show that the rotating equilibrium of Ref.~\onlinecite{Finn_Antonsen_83} can be found within our RxMHD formalism without the need to invoke angular momentum conservation as a constraint.

In contrast to the axisymmetric equilibrium, we showed for the time-dependent waves in Sec.~\ref{sec:WKBMvRxHD} that the solvability condition $\curl(\vec{u}\cross\vec{B}) = 0$ for the potential $\Phi$ in the ideal electrostatic Ohm's law \eqref{eq:OhmPhiLaw} is \emph{not} in general satisfied dynamically. Nevertheless, as the applicability of our relaxed dynamics is limited to very long timescales, it may be reasonable to assume that the electrostatic approximation $\vec{E} = -\grad\Phi$ still holds during dynamical evolution. Then we could build in the ideal Ohm's Law \eqref{eq:OhmPhiLaw} as a holonomic constraint by replacing $\vec{u}$ with $\vec{u}_\perp + u_\parallel\vec{B}/B$, the set of three free fields comprising the components of $\vec{u}$ being replaced by the set of two free fields $\{\Phi, u_\parallel\}$. 

However, adding extra constraints is against the spirit of the relaxation theory we have put forward in this paper, so we advance here the speculation that there may be physical cases where it is not necessary to impose the ideal Ohm's law
through the following heuristic argument: When a plasma is perturbed away from equilibrium, the turbulence level rises to activate relaxation mechanisms. Then a turbulent dynamo effect, \cite{Squire_Bhattacharjee_2016,Moffatt_2014,Yokoi_2013} comes into play, generating an ``anomalous'' e.m.f. such that the ideal Ohm's Law \eqref{eq:OhmPhiLaw} no longer applies. 

This is highly speculative and takes us well beyond the scope and motivation of this paper. As outlined in the Introduction, our main motivation is to develop a tractable and well posed computational method for calculating the slow dynamics of nonaxisymmetric toroidally confined plasmas based on the MRxMHD nested-toroidal-layer model. For this purpose it would suffice to show that, as the number of layers increase (so the depths of the layers decrease) the effect of any violation of the ideal Ohm's Law becomes progressively less significant.

\section{Conclusion}\label{sec:Concl}

We have shown that, unlike the Configuration Space Lagrangian approach,  the Phase Space Lagrangian successfully allows the extension of a dynamical formalism, Relaxed Magnetohydrodynamics (RxMHD), \cite{Dewar_Yoshida_Bhattacharjee_Hudson_2015} to allow coupling between fluid and magnetic field using a cross-helicity constraint. This improves the theoretical basis for dynamical extensions of present Multiregion RxMHD (MRxMHD) equilibrium computations with flow.  \cite{Qu_etal_2020} An axisymmetric steady-flow RxMHD solution is found that meets the consistency test of also being a well-known ideal-MHD rotating equilibrium, but application of the theory to \emph{non}axisymmetric systems is left to future work. 

Also for further work is the development of a linear MRxMHD normal mode code and a nonlinear MRxMHD time evolution code. To maintain the computational advantage exploited in the current SPEC code of computing the magnetic field by solving the Beltrami equation, a simple elliptic PDE, we also need to develop a quasi-adiabatic, slow manifold \cite{Burby_2017} version of the coupled Eqs.~(\ref{eq:AHamPrEL}) and (\ref{eq:xiHamPrELomega1}). In that way $\vec{u}$ would be slaved to $\vec{B}$ and ellipticity restored in the modified Beltrami equation, \eqref{eq:AHamPrEL}.

While the principal motivation for this work is the extension of a computational model, and the formal development has intentionally started from basic Lagrangian and Hamiltonian concepts, our mathematical development includes significant innovation and there is potential for wider physical application.

In particular the noncanonical $\vec{u}$, $\vec{v}$ formalism is very recent, \cite{Burby_2017} and we have taken care here to develop it from first principles and interpret the significance of the seemingly redundant velocity field $\vec{v}$ as the velocity relative to a reference flow. Thus $\vec{u}$ is seen as the net fluid velocity in the Lab frame. This approach may prove useful in other applications, for instance in gyroviscous MHD, \cite{Lingam_Morrison_2014} in solar, space and astrophysics, and in geophysical applications to stratified flows.

We have also developed an elementary discussion of ideal-MHD Eulerian mass, entropy and magnetic flux constraints as a continuous symmetry within the space of feasible evolutions, without calling on the heavy mathematical machinery of geometric MHD mechanics. \cite{Holm_Marsden_Ratiu_Weinstein_85,Holm_Marsden_Ratiu_98}

There is more work to be done in the physical interpretation of our modified MHD. We have demonstrated that linearized perturbations break consistency with the ideal Ohm's Law and suggested how to enforce this by adding an $\vec{E}\cross\vec{B}$ constraint, but have left further analysis to future work. Also connections with turbulent dynamo theory are yet to be explored.

Other mathematical developments might  improve treatment, e.g., of boundary dynamics. \cite{Kats_2001} Also, development of shallow layer versions of MRxMHD would be useful for examining the many-interface limit.

\section*{Appendices}
\appendix

\section{Lie symmetry of Eulerian conservation constraints}\label{sec:Lie}

The variations, Eqs.~(\ref{eq:vvar}--\ref{eq:Bvar}), in the fields that are holonomically constrained to vary with $\bm{\xi}$ can also be written
\begin{align}
    \Delta\vec{v} & =\frac{\d\bm{\xi}}{\d t} \;,\label{eq:vvar2}\\ 
    \Delta\ln\rho &= -\divv\bm{\xi} \label{eq:rhovar2} \\
    \Delta\ln p &= -\gamma\divv\bm{\xi} \label{eq:pvar2} \\
    \Delta\vec{B} & =\vec{B}\dotv(\grad\bm{\xi}  - \Ident\divv\bm{\xi}) \;,\label{eq:Bvar2}
\end{align}
where $\Delta \equiv \delta + \bm{\xi}\dotv\grad$ is the Lagrangian variation operator, \cite{Dewar_70} mentioned in Subsec.~\ref{sec:IMHDmicro} where it was noted that $\Delta\vec{x} = \bm{\xi}$ because $\delta\vec{x} \equiv 0$.

Note also that $\delta$ commutes with $\grad$ and $\partial_{t}$, because Eulerian variations are taken with $\vec{x}$ and $t$ held constant, and
the relations for commuting $\delta$ and $\grad$ with $\d/\d t$ are 
\begin{equation}\label{eq:deltadtgradcommut}
	\delta\frac{\d}{\d t} = \frac{\d}{\d t}\delta +\delta\vec{v}\dotv\grad\;, \:\: \grad\frac{\d}{\d t} = \frac{\d}{\d t}\grad + \grad\vec{v}\dotv\grad \;. 
\end{equation}
Using \eqref{eq:vvar2} and \eqref{eq:deltadtgradcommut} we can now show that the \emph{Lagrangian variation $\Delta$ and the advective derivative ${\d}/{\d t}$ commute},
\begin{align}\label{eq:Deltadtcommut}
	\Delta\frac{\d}{\d t} &= \frac{\d}{\d t}\delta + \bm{\xi}\dotv\grad\frac{\d}{\d t} + \delta\vec{v}\dotv\grad \nonumber\\
			&= \frac{\d}{\d t}\delta + \bm{\xi}\dotv\frac{\d}{\d t}\grad + \bm{\xi}\dotv(\grad\vec{v})\dotv\grad + \delta\vec{v}\dotv\grad \nonumber\\
			&= \frac{\d}{\d t}\delta + \frac{\d}{\d t}\bm{\xi}\dotv\grad - \frac{\d\bm{\xi}}{\d t}\dotv\grad + \Delta\vec{v}\dotv\grad \nonumber\\
			&= \frac{\d}{\d t}\Delta - \frac{\d\bm{\xi}}{\d t}\dotv\grad + \Delta\vec{v}\dotv\grad \nonumber\\
			&= \frac{\d}{\d t}\Delta \;,
\end{align}

The relation for commuting $\Delta$ and $\grad$ is
\begin{align}\label{eq:Deltagradcommut}
	\Delta\grad &= \grad\delta + \bm{\xi}\dotv\grad\grad \nonumber\\ 
	&= \grad(\delta + \bm{\xi}\dotv\grad) - (\grad\bm{\xi})\dotv\grad \nonumber\\
	&= \grad\Delta - (\grad\bm{\xi})\dotv\grad \;. 
\end{align}

As Eqs.~(\ref{eq:Continuity}) and (\ref{eq:rhovar2}) are of the same form as Eqs.~(\ref{eq:AdiabaticPressure} and (\ref{eq:pvar2}), with $p \mapsto \rho$ and $\gamma \mapsto 1$, we need consider only the pressure constraint PDE, \eqref{eq:AdiabaticPressure}, without loss of generality. Collecting all terms of the advective form of \eqref{eq:AdiabaticPressure} to its LHS, dividing by $p$, and applying $\Delta$ to the new LHS gives, using \eqref{eq:vvar2}, \eqref{eq:pvar2}, and Eqs.~(\ref{eq:deltadtgradcommut}--\ref{eq:Deltagradcommut}), 
\begin{align}\label{eq:DeltaAdiabaticPressure}
	&\Delta\left(\frac{\d \ln p}{\d t}  + \gamma \divv\vec{v}\right) \nonumber\\
	\quad&= \frac{\d}{\d t}\Delta\ln p  + \gamma \divv\Delta\vec{v} - \gamma[(\grad\bm{\xi}) \dotv\grad]\dotv\vec{v} \nonumber\\
	\quad&= -\gamma\frac{\d}{\d t}\divv\bm{\xi}  + \gamma \divv\frac{\d \bm{\xi}}{\d t}  - \gamma (\grad\bm{\xi}) \ddotv\grad\vec{v} \nonumber\\
	\quad&= -\gamma\frac{\d}{\d t}\divv\bm{\xi}  + \gamma \left(\frac{\d}{\d t}\grad + \grad\vec{v}\dotv\grad\right)\dotv\bm{\xi}  - \gamma (\grad\bm{\xi}) \ddotv\grad\vec{v} \nonumber\\
	\quad&=\gamma (\grad\vec{v})\ddotv\grad\bm{\xi}  - \gamma (\grad\bm{\xi}) \ddotv\grad\vec{v} \nonumber\\
	\quad&= 0 \;,
\end{align}
thus verifying that the Eulerian holonomic variations Eqs.~(\ref{eq:vvar}) and \eqref{eq:pvar} make an infinitesimal transformation to a new solution of \eqref{eq:AdiabaticPressure}. $\Box$ (In the above we used the double dot product of two dyadics, say $\vsf{a}$ and $\vsf{b}$, defined as $\vsf{a}\ddotv\vsf{b} \equiv \sum_{i,j}a_{i,j}b_{j,i} = \sum_{i,j}b_{j,i}a_{i,j} \equiv \vsf{b}\ddotv\vsf{a}$.)

It remains to verify that the flux-freezing condition \eqref{eq:FrozenFlux} is also preserved under the holonomic variations Eqs.~(\ref{eq:vvar}) and \eqref{eq:Bvar}. Collecting all terms of the advective form of \eqref{eq:FrozenFlux} to its LHS and applying $\Delta$ gives, using \eqref{eq:vvar2}, \eqref{eq:Bvar2}, and Eqs.~(\ref{eq:deltadtgradcommut}--\ref{eq:Deltagradcommut}),
\begin{align}
	&\Delta\left[\frac{\d \vec{B}}{\d t} + \vec{B}\dotv(\Ident\divv\vec{v} - \grad\vec{v})\right] \nonumber\\ %
	&= \frac{\d}{\d t}\Delta\vec{B} + (\Delta\vec{B})\dotv(\Ident\divv\vec{v} - \grad\vec{v}) \nonumber\\
	&\quad + \vec{B}[\grad\Delta - (\grad\bm{\xi})\dotv\grad]\dotv\vec{v} - \vec{B}\dotv [\grad\Delta - (\grad\bm{\xi})\dotv\grad]\vec{v} \nonumber\\ 
	&= \frac{\d}{\d t}[\vec{B}\dotv(\grad\bm{\xi}  - \Ident\divv\bm{\xi})] + \vec{B}\dotv(\grad\bm{\xi}  - \Ident\divv\bm{\xi})\dotv(\Ident\divv\vec{v} - \grad\vec{v}) \nonumber\\
	&\quad + \vec{B}\divv\frac{\d\bm{\xi}}{\d t} - \vec{B}(\grad\bm{\xi})\ddotv(\grad\vec{v}) 
	- \vec{B}\dotv\grad\frac{\d\bm{\xi}}{\d t} + \vec{B}\dotv(\grad\bm{\xi})\dotv\grad\vec{v} \nonumber\\ 
	&= \frac{\d \vec{B}}{\d t}\dotv(\grad\bm{\xi}  - \Ident\divv\bm{\xi}) + \vec{B}\dotv(\frac{\d}{\d t}\grad)\bm{\xi}  - \vec{B}(\frac{\d}{\d t}\grad)\dotv\bm{\xi} \nonumber\\
	&\quad + \vec{B}\dotv(\grad\bm{\xi}  - \Ident\divv\bm{\xi})\dotv(\Ident\divv\vec{v} - \grad\vec{v}) \nonumber\\
	&\quad + \vec{B}\divv\frac{\d\bm{\xi}}{\d t} - \vec{B}(\grad\bm{\xi})\ddotv(\grad\vec{v}) 
	- \vec{B}\dotv\grad\frac{\d\bm{\xi}}{\d t} + \vec{B}\dotv(\grad\bm{\xi})\dotv\grad\vec{v} \nonumber\\ 
	&= \frac{\d \vec{B}}{\d t}\dotv(\grad\bm{\xi}  - \Ident\divv\bm{\xi}) + \vec{B}\dotv(\grad\bm{\xi}  - \Ident\divv\bm{\xi})\dotv(\Ident\divv\vec{v} - \grad\vec{v}) \nonumber\\
	&\quad + \vec{B}\dotv\left[\grad\frac{\d}{\d t} - (\grad\vec{v})\dotv\grad\right]\bm{\xi}  - \vec{B}\left[\grad\frac{\d}{\d t} - (\grad\vec{v})\dotv\grad\right]\dotv\bm{\xi} \nonumber\\
	&\quad + \vec{B}\divv\frac{\d\bm{\xi}}{\d t} - \vec{B}(\grad\bm{\xi})\ddotv(\grad\vec{v}) 
	- \vec{B}\dotv\grad\frac{\d\bm{\xi}}{\d t} + \vec{B}\dotv(\grad\bm{\xi})\dotv\grad\vec{v} \;. \nonumber
\end{align}
	
Now cancel all four $\d\bm{\xi}/\d t$ terms:
\begin{align}
	&\Delta\left[\frac{\d \vec{B}}{\d t} + \vec{B}\dotv(\Ident\divv\vec{v} - \grad\vec{v})\right] \nonumber\\ %
	&= \frac{\d \vec{B}}{\d t}\dotv(\grad\bm{\xi}  - \Ident\divv\bm{\xi}) + \vec{B}\dotv(\grad\bm{\xi}  - \Ident\divv\bm{\xi})\dotv(\Ident\divv\vec{v} - \grad\vec{v}) \nonumber\\
	&\quad - \vec{B}\dotv(\grad\vec{v})\dotv\grad\bm{\xi}  + \vec{B}\dotv(\grad\bm{\xi})\dotv\grad\vec{v} \nonumber\\
	&\quad + \vec{B}(\grad\vec{v})\ddotv(\grad\bm{\xi}) - \vec{B}(\grad\bm{\xi})\ddotv(\grad\vec{v}) \nonumber\\ 
	&= \frac{\d \vec{B}}{\d t}\dotv(\grad\bm{\xi}  - \Ident\divv\bm{\xi}) - \vec{B}(\divv\bm{\xi})\divv\vec{v} + \vec{B}\dotv(\grad\vec{v})\divv\bm{\xi} \nonumber\\
	&\quad + \vec{B}\dotv(\grad\bm{\xi})\divv\vec{v} - \underline{\vec{B}\dotv(\grad\bm{\xi})\dotv\grad\vec{v}}  \nonumber\\
	&\quad - \vec{B}\dotv(\grad\vec{v})\dotv\grad\bm{\xi}  + \underline{\vec{B}\dotv(\grad\bm{\xi})\dotv\grad\vec{v}} \nonumber\\ 
	&= \left[\frac{\d \vec{B}}{\d t}+ \vec{B}\dotv(\Ident\divv\vec{v} - \grad\vec{v})\right]\dotv(\grad\bm{\xi}  - \Ident\divv\bm{\xi}) \;,   \label{eq:deltaFrozenFlux}
\end{align}
which vanishes provided \eqref{eq:FrozenFlux} is satisfied on the unvaried evolution. That is, if the original evolution is a solution of \eqref{eq:FrozenFlux} then so is the varied evolution.

Summarizing, in this Appendix we have shown that $\bm{\xi}$ generates an infinitesimal mapping of the space of feasible evolutions onto itself through Newcomb's holonomic constraints, Eqs.~(\ref{eq:vvar}--\ref{eq:Bvar}).

\section{Energy principle for fully relaxed equilibria with flow}\label{sec:genTaylor}

In this section we extend Taylor's energy minimization principle by including flow and thermal kinetic energies and keeping Taylor's magnetic helicity constraint, adding an entropy constraint, and adding the cross helicity constraint in order to construct a relaxed state with finite pressure and a steady flow, in a similar way to Finn and Antonsen. \cite{Finn_Antonsen_83}  We take as energy the Hamiltonian $H_{\rm nc}$, \eqref{eq:Hstd}. 

To implement the relaxation prescription in Sec.~\ref{sec:Relax}, the global invariants listed in Sec.~\ref{sec:FIhel} are enforced using Lagrange multipliers to give the constrained energy functional
\begin{equation}\label{eq:HRx}
\begin{split}
	H_\Omega^{\rm Rx} [\rho, \vec{u}, p, \vec{A}]
	&\equiv  H_\Omega  [\rho, \vec{u}, p, \vec{A}] \\
	&- \tau_\Omega  S_\Omega  - \mu_\Omega  K_\Omega 
	 - \XHelmult  K_\Omega ^{\rm X}[\vec{u},\vec{A}] 
	\, , 
\end{split}
\end{equation}
where $\tau_\Omega $, $\mu_\Omega $, and $\XHelmult $ are Lagrange multipliers to enforce conservation, respectively, of entropy, magnetic helicity, and cross helicity in $\Omega$. 

The energy variation is now, assuming the support of each variation is localized within $\Omega$,
\begin{equation}\label{eq:WRxvar}
\begin{split}
	\delta H_\Omega^{\rm Rx} &=  \int_\Omega \left(\delta\vec{u}\dotv\frac{\delta H_\Omega}{\delta\vec{u}}
	+ \delta\vec{A}\dotv\frac{\delta H_\Omega}{\delta\vec{A}} \right.\\
	&\left. \quad + \,\delta p\frac{\delta H_\Omega}{\delta p}
	+ \rho\bm{\xi}\dotv\grad\frac{\delta H_\Omega}{\delta\rho} \right)\d V \\ 
	&\quad -\tau_\Omega \,\delta S_\Omega  - \mu_\Omega \, \delta K_\Omega  - \XHelmult \, \delta K_\Omega ^{\rm X} 
	\, . 
\end{split}
\end{equation}
The functional derivatives are
\begin{align}
	\frac{\delta H_\Omega^{\rm Rx} }{\delta\vec{u}} &= \rho\vec{u} 
	- \XHelmult \frac{\vec{B}}{\muSI} 
	\;,\label{eq:uvarWRx}\\
	\frac{\delta H_\Omega^{\rm Rx} }{\delta \vec{A}}  &= \frac{1}{\muSI}\left(\curl\vec{B} - \mu_\Omega \vec{B}
	- \XHelmult  \curl\vec{u}\right) 
	\;,\label{eq:AvarWRx} \\
	\frac{\delta H_\Omega^{\rm Rx} }{\delta p}  &= \frac{1}{\gamma-1}\left(1 - \tau_\Omega\frac{\rho}{p}\right) \;,\label{eq:pvarWRx}\\
	\frac{\delta H_\Omega^{\rm Rx} }{\delta\rho}  &=   \frac{u^2}{2} 
	 	 -\frac{\tau_\Omega }{\gamma - 1}\left[\ln\left(\kappa\frac{p}{\rho^\gamma}\right) 
		  -\gamma\right] \label{eq:rhovarWRx} \;.
\end{align}

For $\delta H_\Omega^{\rm Rx}$ to be zero for independent variations $\delta\vec{u}$, $\delta\vec{A}$, $\delta p$, and $\bm\xi$ the four Euler--Lagrange equations
\begin{align}
	\rho\vec{u} &= \XHelmult \frac{\vec{B}}{\muSI}  \;,\label{eq:uEnPrEL} \\
	\curl\vec{B} &= \mu_\Omega  \vec{B}  
					+ \XHelmult\curl\vec{u}  \;,\label{eq:AEnPrEL}\\
			p &= \tau_\Omega  \rho \label{eq:pEnPrEL}\;, \\
	\text{and}\:\: \grad h_\Omega &= 0 \label{eq:xiEnPrEL}
\end{align}
must be satisfied. In the above we have denoted $\delta H_\Omega^{\rm Rx}/\delta\rho$ by $h_\Omega$, the Bernoulli head, defined by
\begin{equation}\label{eq:Head}
\begin{split}
	h_\Omega &\equiv \frac{u^2}{2} 
	 	 -\frac{\tau_\Omega }{\gamma - 1}\left[\ln\left(\kappa\frac{p}{\rho^\gamma}\right) 
		  -\gamma\right] + \const\\
		  &= \frac{u^2}{2} + \tau_\Omega \ln\frac{\rho}{\rho_\Omega} \;,
\end{split}
\end{equation}
where the second line absorbs the arbitrary constant in the definition into $\rho_\Omega$. 
Equation (\ref{eq:xiEnPrEL}) shows $h_\Omega$ is constant throughout $\Omega$ in relaxed steady flow. Choosing this constant to be \emph{zero}, gives us an expression for the physical observable $\rho$, found from \eqref{eq:Head} to be given by
\begin{equation}\label{eq:rhoBoltz}
\begin{split}
	\rho &= \rho_\Omega\exp\left(-\frac{u^2}{2\tau_\Omega}\right) \;.
\end{split}
\end{equation}

In the limit $\vec{u} = 0$, $\rho$ and hence $p$ are constant within $\Omega$, as was assumed in previous MRxMHD work, e.g. in developing the Stepped Pressure Equilibrium Code SPEC. \cite{Hudson_etal_2012b}

Note that \eqref{eq:xiEnPrEL} can also be written, using \eqref{eq:uEnPrEL} and \eqref{eq:AEnPrEL} and the identity $\grad(u^2/2) = \vec{u}\dotv\grad\vec{u} -(\curl\vec{u})\cross\vec{u}$, 
\begin{equation}\label{eq:AltxiEnPrEL}
\begin{split}
	\rho\vec{u}\dotv\grad\vec{u} 
	&= -\grad p + \vec{j}\cross\vec{B} \;,
\end{split}
\end{equation}
where $\vec{j} \equiv \curl\vec{B}/\muSI$ by Amp\`ere's Law (pre Maxwell).

\subsection{Equilibrium Consistency checks}\label{sec:FldAlgnedFloConsistency}

Below we show Eqs.~(\ref{eq:uEnPrEL}--\ref{eq:xiEnPrEL}) are IMHD-equilibrium compatible, i.e. consistent with Eqs.~(\ref{eq:Continuity}--\ref {eq:idealEqMot}), $\vec{v}$ there being replaced by $\vec{u}$ and terms in $\partial_t$ being set to zero: 

\begin{enumerate}
\item Observing that \eqref{eq:AEnPrEL} is compatible with $\divv\vec{B} = 0$, take the divergence of both sides of \eqref{eq:uEnPrEL}, to find $\divv(\rho\vec{u}) = 0$, thus showing $\vec{u}$ satisfies the continuity equation \eqref{eq:Continuity}, at least in the steady flow case. $\Box$
\item From \eqref {eq:uEnPrEL} we have $\vec{u}\cross\vec{B} = 0$, so the static ideal Ohm's law, \eqref{eq:OhmPhiLaw} is trivially satisfied under the ergodic relaxation condition $\grad\Phi = 0$. $\Box$
\item The static limit of the ideal equation of motion, \eqref{eq:idealEqMot}, is just \eqref{eq:AltxiEnPrEL}. $\Box$
\item The final consistency test must be more nuanced, as we have altered the ideal thermodynamics by relaxing the temperature (proportional to $\tau$) throughout $\Omega$. This models the rapid transport of heat in the highly chaotic magnetic field line flow that is required to justify relaxation theory physically, but it has the consequence that $p/\rho^\gamma$ is not constant microscopically (though $p V_\Omega ^\gamma$ is still constant in time if $p$ is spatially constant). This switch to a local isothermal equation of state can be modeled by taking $\gamma \to 1$ so the static version of \eqref{eq:AdiabaticPressure} is of the same form as \eqref {eq:rhovar}, i.e. $\divv(p\vec{u}) = \tau_\Omega \divv(\rho\vec{u}) = 0$, which is satisfied because $\divv(\rho\vec{u}) = 0$ was verified in the first of our consistency tests. $\Box$
\end{enumerate}
(We use the notation $\Box$ to indicate that an equation has passed a specified validation test.) 

Interestingly,  Eqs.~(\ref{eq:uEnPrEL}--\ref{eq:xiEnPrEL}) would also be compatible with steady, relaxed \emph{Euler} flow, \cite{Sato_Dewar_2017} if we could suppose $\vec{B}$ is a harmonic ``vacuum'' field, so that $\curl\vec{B} = 0$. Then $\vec{u}$ would obey the nonlinear Beltrami equation $\curl\vec{u} = -(\muSI\mu_\Omega/\XHelmult^2)\rho\vec{u}$ if we could also satisfy $\curl(\rho\vec{u}) = 0$.

\section{Axisymmetric equilibria with both field-aligned and cross-field flow}\label{sec:RxMHDsteady}

The first test of our new phase-space action formulation is whether it can generalize the rather restricted class of flows in the relaxed equilibria derived from an energy principle in Sec.~\ref{sec:genTaylor}, in which $\vec{u}$ had to be parallel to $\vec{B}$. As seen from \eqref{eq:uHamPrEL}, the new field $\vec{v}$ does indeed provide the possibility of flows with a component perpendicular to $\vec{B}$, even in steady flows, in which it is also appropriate to check for consistency with IMHD. We term such equilibria ``semi-relaxed'' as they do not obey the ergodic relaxation condition $\vec{u}\cross\vec{B} = 0$ discussed in Sec.~\ref{sec:IOhm}.

Previous authors \cite{Woltjer_58b, Dennis_Hudson_Dewar_Hole_14a} have inserted toroidal equilibrium flow ``by hand''  by constraining the $Z$ component of the angular momentum in their energy stationarization (or, equivalently, entropy stationarization \cite{Finn_Antonsen_83}). However, this is physically consistent only if the system is rotationally symmetric about the $Z$ axis. In our more general approach, conserved physical quantities should arise naturally from Noether's theorem if there is a continuous symmetry, rather than by constraints. 

\subsection{Rigid rotation v}\label{sec:FAcomparison}

Nevertheless the previous work on axisymmetric equilibria does provide physically consistent solutions that can be used to validate our equations, so we now test whether we can reproduce the results of Finn and Antonsen (FA) \cite{Finn_Antonsen_83} on relaxed axisymmetric equilibria. 

In the following discussion we use the usual cylindrical coordinates $R, \phi, Z$, with $R$ the distance from the vertical ($Z$) axis and $\phi$ the toroidal angle, $\esub{R,\phi,Z}$ being the corresponding right-handed orthonormal set of basis vectors.

Specifically, we seek to show that choosing $\vec{v}$ to be a rigid rotation around the $Z$-axis with angular frequency $\varpi_\Omega$,
\begin{equation}\label{eq:vFA}
	\vec{v} = R\varpi_\Omega \esub{\phi}(\phi) = R^2\varpi_\Omega \grad{\phi} \;,
\end{equation}
throughout an axisymmetric toroidal domain $\Omega$, will lead to a time-independent solution of the RxMHD equations (\ref{eq:uHamPrEL}--\ref{eq:xiHamPrELomega1}) that is consistent with FA's Eqs.~(26), (27), and (29). \cite{Finn_Antonsen_83} Transcribed into our notation the FA equations are 
\begin{align}
	&\vec{u} = 
	 u_\parallel^{\rm Rx}\esub{\parallel} + R\varpi_\Omega \esub{\phi}(\phi) \label{eq:FA26}\\
	&\curl\left[\left(1 - M^{\rm Rx\,2}_{\rm A}\right)\vec{B}\right] 
	= \mu_\Omega \vec{B} + 2\XHelmult \varpi_\Omega \esub{Z} \label{eq:FA27} \\
	& \frac{u_\parallel^{{\rm Rx}\, 2}}{2} - \frac{R^2\varpi_\Omega^2}{2}
	 + \tau_\Omega\ln\frac{\rho}{\rho_\Omega} = 0 \label{eq:FA29} \;.
\end{align}
where $u_\parallel^{\rm Rx}\equiv \XHelmult B/\muSI\rho$ is the fully relaxed flow speed defined in \eqref{eq:ufullRx}, and 
\begin{equation}\label{eq:MAdef}
	M^{\rm Rx}_{\rm A} \equiv \frac{u_\parallel^{\rm Rx}}{c_{\rm A} }
\end{equation}
is the \emph{parallel Alfv\'en Mach number}, $c_{\rm A} \equiv (B^{2}/\muSI \rho)^{1/2}$ being the local Alfv\'en speed as in Sec.~\ref{sec:WKBIMHD}.

With the choice \eqref{eq:vFA}, we note first that \eqref{eq:FA26} and \eqref{eq:uHamPrEL} are identical.  $\Box$ 

Also, multiplying both sides of \eqref{eq:FA26} by $\rho$ and taking divergences gives
\begin{equation}\label{eq:uHamPrELdiv}
\begin{split}
	\divv(\rho\vec{u})
	 &= \varpi_\Omega R^2\rho\divv\grad\phi + \varpi_\Omega \grad\phi\dotv \grad(R^2\rho) = 0
	\;,
\end{split}
\end{equation}
thus verifying consistency with the continuity equation, \eqref{eq:ucont}. \!\!\!$\Box$ [In deriving the above identities we have used $\grad\phi = \esub{\phi}/R$, $\nabla^2\phi = 0$, and $\esub{\phi}\dotv\grad (R^2\rho) = 0$, as $\rho = \rho(R,Z)$ by axisymmetry.]

The vorticity of the toroidal flow is a constant vector in the $Z$ direction, $\curl\vec{v} = 2\varpi_\Omega\esub{Z}$. We can also calculate the total vorticity, $\bm{\omega} \equiv \curl\vec{u}$, by taking the curl of both sides of \eqref{eq:FA26},
\begin{equation}\label{eq:omegaFA}
\begin{split}
	\bm{\omega} &= \curl\left(\frac{\XHelmult \vec{B}}{\muSI\rho}\right) + 2\varpi_\Omega\esub{Z} \\
	&= \left(\frac{\XHelmult}{\muSI\rho}\right)\curl\vec{B} + \grad\!\left(\frac{\XHelmult}{\muSI\rho}\right)\!\cross\vec{B}
	+ 2\varpi_\Omega\esub{Z}	 \;.
\end{split}
\end{equation}
With this identification the first line of \eqref{eq:FA27} can now, with a little rearranging and multiplying both sides by $\muSI$, be recognized as \eqref{eq:AHamPrEL}, $\curl\vec{B} = \mu_\Omega \vec{B} +\XHelmult \bm{\omega}$. $\Box$

It is convenient at this point to introduce a poloidal-toroidal decomposition, i.e. we use two basis vectors spanning the poloidal, $R,Z$ half-plane at each toroidal angle $\phi$, and a third basis vector in the orthogonal toroidal direction $\esub{\phi}(\phi)$. The most general representation of $\vec{B}$ (and similarly other divergence-free fields like $\curl\vec{B}$, $\rho\vec{u}$, and $\bm{\omega}$) is then 
\begin{equation}\label{eq:genGSrepB}
	\vec{B} = \grad\phi\cross\grad\psi + F \grad \phi \;,
\end{equation}
the first term on the RHS being the \emph{poloidal magnetic field}, $\vec{B}_{\rm pol}$ and the second the \emph{toroidal magnetic field}, $\vec{B}_{\rm tor}$. Unlike in the Grad--Shafranov representation for flowless MHD equilibria (see e.g.  Ref.~\onlinecite[pp. 177--178]{Hosking_Dewar_2015}), the toroidal field strength function $F$ is not simply a function of $\psi$---at this point it is an unspecified function of $R$ and $Z$. 

Taking the curl of both sides gives, in the expected toroidal-poloidal form,
\begin{equation}\label{eq:curlBgenGS}
\begin{split}
	\curl \vec{B} 
	&= -\grad\phi\cross\grad F + \curl\left(\grad\phi\cross\grad\psi\right) \\
	&= -\grad\phi\cross\grad F + \left[\nabla^2\psi\, \grad\phi \right. \\
	&\quad\quad \quad \quad +\left.(\grad\grad\phi)\dotv\grad\psi - (\grad\phi)\dotv \grad\grad\psi\right] \\
	&=  -\grad\phi\cross\grad F + \Delta^*\psi\, \grad\phi \;,
\end{split}
\end{equation}
where,  using the identity $\grad\psi\dotv\grad\grad\phi - \grad\phi\dotv\grad\grad\psi = -(2/R)(\grad R\dotv\grad\psi)\grad\phi$, Ref.~\onlinecite[pp. 177--178]{Hosking_Dewar_2015}, $\Delta^*\psi \equiv R^2\divv(\grad\psi/R^2)$.

Multiplying both sides of \eqref{eq:FA26} by $\rho$ and using \eqref{eq:genGSrepB} we find 
\begin{equation}\label{eq:genGSreprhou}
\begin{split}
	\rho\vec{u} 
	&= \grad\phi\cross\grad\left(\frac{\XHelmult \psi}{\muSI}\right) 
	+ \left(\frac{\XHelmult}{\muSI}F + \varpi_\Omega R^2 \rho \right)\grad\phi 
	\;,
\end{split}
\end{equation}
which again is in the expected poloidal-toroidal representation. (Note that the poloidal flow is driven solely by cross helicity---setting $\XHelmult = 0$ gives a purely toroidal, rigid-rotational flow.)

Substituting \eqref{eq:genGSrepB} and \eqref{eq:curlBgenGS} in \eqref{eq:AHamPrEL} yields a toroidal-poloidal form for the vorticity,
\begin{align}
	\bm{\omega} &= \frac{1}{\XHelmult}\curl\vec{B} - \frac{\mu_\Omega}{\XHelmult}  \vec{B}  \label{eq:genGSvorticity}\\
				&= - \grad\phi\cross \grad\left(\frac{F + \mu_\Omega\psi}{\XHelmult} \right)  
				+ \left(\frac{\Delta^*\psi - \mu_\Omega F}{\XHelmult}\right)\grad \phi  \;,\nonumber
\end{align}
whereas taking the curl of both sides of \eqref{eq:FA26} 
gives an alternative expression for the vorticity,
\begin{equation}\label{eq:genGSrepvort}
\begin{split}
	\bm{\omega} 
	&= \frac{\XHelmult}{\muSI\rho} \left(\curl\vec{B} - \frac{\grad\rho}{\rho}\cross\vec{B}\right) + \curl\vec{v}\\	
	&= -\grad\phi\cross\grad\left(\frac{\XHelmult}{\muSI}\frac{F}{\rho} + \varpi_\Omega R^2 \right) \\ 
	&\quad + \frac{\XHelmult}{\muSI\rho} \left(\Delta^*\psi 
	- \frac{\grad\rho\dotv\grad\psi}{\rho}\right)\grad\phi  \;.
\end{split}
\end{equation}

Substituting \eqref{eq:vFA} and \eqref{eq:genGSrepvort} in the force-balance equation \eqref{eq:xiHamPrELomega1} (with $\partial_t\vec{u} = 0$) we get
\begin{equation}\label{eq:genGSforcebal}
\begin{split}
	\grad\left[ h_\Omega-\varpi_\Omega\left(\frac{\XHelmult}{\muSI}\frac{F}{\rho} + \varpi_\Omega R^2 \right)\right] &= 0 
	\;,
\end{split}
\end{equation}
thus generalizing the Bernoulli relation \eqref{eq:xiEnPrEL} to include rigid rotation. Choosing the arbitrary constant $\rho_\Omega$ appropriately, \eqref{eq:genGSforcebal} implies
\begin{equation}\label{eq:genBern3}
	\frac{u^2}{2} + \tau_\Omega\ln\frac{\rho}{\rho_\Omega} - \frac{\XHelmult\varpi_\Omega}{\muSI}\frac{F}{\rho} - \varpi_\Omega^2 R^2  = 0 \;.
\end{equation}
Then, expanding $u^2/2$ using \eqref{eq:FA26}, \eqref{eq:genBern3} is readily seen to agree with the generalized Bernoulli equation, \eqref{eq:FA29}. $\Box$

Another form for the generalized Bernoulli equation may be had by decomposing \eqref{eq:FA26} into poloidal and toroidal components,
$\vec{u}_\theta \equiv \XHelmult\esub{\phi}\cross\grad\psi/R\muSI\rho$ and $u_\phi \equiv \XHelmult F/R\muSI\rho + R\varpi_\Omega$, respectively. Then $u^2 = u_\theta^2 + u_\phi^2$ and $\XHelmult\varpi_\Omega F/\muSI\rho\, + \varpi_\Omega^2 R^2 = \varpi_\Omega R u_\phi $. Thus \eqref{eq:genBern3} can also be written
\begin{equation}\label{eq:genBern4}
	\tau_\Omega\ln\frac{\rho}{\rho_\Omega} + \frac{u_\theta^2 + u_\phi^2}{2} - \varpi_\Omega R u_\phi   = 0 \;,
\end{equation}
as in the IMHD result of McClements and Hole, Ref.~\onlinecite[Eq. (20)]{McClements_Hole_10}, in the case of isothermal magnetic surfaces (identifying their $2T/m_i$ with our $\tau_\Omega$). This reference discusses the conditions under which the Grad--Shafranov equation, to be derived in the next subsection, changes from elliptic to hyperbolic.

Finally, crossing both sides of \eqref{eq:FA26} with $\vec{B}$ and using \eqref{eq:genGSrepB} we find
\begin{equation}\label{eq:OhmConstraint}
\begin{split}
	\vec{u}\cross\vec{B} &= \varpi_\Omega R^2\grad\phi\cross\vec{B} = -\varpi_\Omega \grad\psi \\
					&= \grad\Phi \;,
\end{split}
\end{equation}
where $\Phi = -\varpi_\Omega\psi$. Comparing with \eqref{eq:OhmPhiLaw} we see that, although we did not impose the ideal Ohm's law as a constraint, it is nevertheless satisfied in this case. $\Box$

\subsection{Grad--Shafranov Equation}\label{sec:GSeq}

Comparing the two expressions for $\bm{\omega}$ given in \eqref {eq:genGSvorticity} and \eqref {eq:genGSrepvort}, and choosing the arbitrary baseline for $\psi$ appropriately, we find two relations
\begin{align}
	\left(1 - M^{\rm Rx\,2}_{\rm A}\right) F 
	&=  \XHelmult\varpi_\Omega R^2 - \mu_\Omega\psi \label{eq:GSB1} \\
	\left(1 - M^{\rm Rx\,2}_{\rm A}\right)\Delta^*\psi &= \mu_\Omega F
	- \frac{\XHelmult^2}{\muSI\rho} \frac{\grad\rho\dotv\grad\psi}{\rho} \;, \label{eq:GSB2}
\end{align}
the first giving $F$ in terms of $\psi$ and $\rho$ and the second being what we shall call a generalized Grad--Shafranov--Beltrami (GSB) equation for $\psi$. (A similar system was analyzed in Ref.~\onlinecite{Dewar_Hudson_Bhattacharjee_Yoshida_2017}, in the limit $\XHelmult \to 0$, $\varpi_\Omega \to 0$ and rippled slab geometry.) 

To solve \eqref{eq:GSB2} we need to express $\rho$ in terms of $\psi$ and $R$, which can be done by using the generalized Bernoulli equation \eqref{eq:FA29}, to give an implicit equation for $\rho$,
\begin{equation}\label{eq:rhoBoltzmannB}
	\rho \equiv \rho_\Omega \exp
	\left[
	-\frac{1}{\tau_\Omega}\left(\frac{\XHelmult ^2}{\muSI\rho}\frac{B^2}{2 \muSI\rho} - \frac{R^2\varpi_\Omega^2}{2}\right)
	\right] \;,
\end{equation}
where $B^2 = (|\grad\psi|^2 + F^2)/R^2$.

There is an obvious singularity in \eqref{eq:GSB2} when the fully-relaxed-flow Alfv\'en Mach number $M^{\rm Rx}_{\rm A} = 1$, but plasma flows in toroidal confinement experiments are typically much less than the Alfv\'en speed, defined using the total magnetic field in the numerator, so it is unlikely this singularity would be encountered in practice. 

However, while not immediately obvious, the dependence of $\rho$ on $|\grad\psi|$, through $B^2$ in \eqref{eq:rhoBoltzmannB}, can cause \eqref{eq:GSB2} to become hyperbolic at much lower plasma flow speeds than $c_{\rm A}$, as first found by Lovelace \emph{et al.} \cite{Lovelace_etal_86} and analyzed in the context of modern low-aspect-ratio tokamaks by McClements and Hole. \cite{McClements_Hole_10} This is because the $\grad\rho$ in \eqref{eq:GSB2} contributes a factor, $\grad\grad\psi$, having second-order derivatives of $\psi$ that must be included along with the second-order derivatives in $\Delta^*\psi$ to evaluate whether \eqref{eq:GSB2} is elliptic or hyperbolic. (This is based on the sign of the discriminant $D = A_{RZ}^2 - 4 A_{RR} A_{ZZ}$, with
$A_{RR}, A_{RZ}, A_{ZZ}$ the coefficients of $\partial^2\psi/\partial^2 R,\, \partial^2\psi/\partial R\partial Z,\, \partial^2\psi/\partial^2 Z$, respectively: If $D=0$ the equation is parabolic, while  $D<0$ implies ellipticity and 
$\D>0$ hyperbolicity.)

It seems unlikely that the MHD equilibria studied in this Appendix would be minima of the Hamiltonian $H_\Omega^{\rm Rx}$ in ranges where such transitions occur, 
which has been confirmed by Hameiri. \cite{Hameiri_98} 
In these cases, while valid MHD equilibria, they could not properly be called relaxed. 

\section{Interface Euler--Lagrange equation}\label{sec:RxMHDELsurf}

To calculate the boundary contribution to the variation of the phase-space action $\Sph \equiv \int\!\d t L_\Omega^{\rm Rx}$, we need to take into account surface terms from integrations by parts omitted in Sec.~\ref{sec:RxMHDEL} because the support of $\bm{\xi}$ was taken not to include $\partial\Omega$. These integrations by parts do not give any surface terms involving $\delta\vec{u}$, 
but we need to include the surface term from the variation of the boundary itself, $\int\!\d t\!\int_{\partial\Omega}\!\d\vec{S} \dotv\bm{\xi}\,\mathcal{L}_\Omega^{\rm Rx}$, where $\mathcal{L}_\Omega^{\rm Rx}$ is given in \eqref{eq:RxPSL}. Using Eqs.~(\ref{eq:vvar}), (\ref{eq:rhovar}), and (\ref{eq:rhovarWRx}), 

in \eqref{eq:Lphx}, we now calculate the residual, boundary action variation 
\begin{align}
	&\delta\Sph = \int\!\d t\!\!\int_\Omega\!\d V\left[\partial_t(\rho\,\vec{u}\dotv\bm{\xi})
	+ \divv(\rho\vec{v}\vec{u}\dotv\bm{\xi} - \rho\,\bm{\xi}\vec{v}\dotv\vec{u})\right]  \nonumber\\
	&+ \int\!\!\d t\!\int_\Omega\!\d V \divv\left\{\rho\,\bm{\xi}\left(\frac{u^2}{2} 
	 	 -\frac{\tau_\Omega }{\gamma - 1}\left[\ln\left(\kappa\frac{p}{\rho^\gamma}\right) 
		  -\gamma\right] \right)\right\}  \nonumber\\
	&-\frac{1}{\muSI}\int\!\!\d t\!\int_\Omega\d V \divv\left[\delta\vec{A}\cross\!\left(\vec{B} 
	 - \frac{\mu_\Omega}{2} \vec{A} - \XHelmult  \vec{u}\right)\right] \nonumber\\
	 &+ \!\int\!\d t\!\int_{\partial\Omega}\!\d\vec{S} \dotv\bm{\xi}\,\mathcal{L}_\Omega^{\rm Rx} 
	 \label{eq:deltaSphraw} \;.
\end{align}

To commute $\int_\Omega\!\d V$ and $\partial_t$ in the first term of the top line (which arose from $\rho\vec{u}\dotv\delta\vec{v})$ we write
\begin{equation*}
\begin{split}
	&\int\!\d t\!\!\int_\Omega\!\d V\partial_t(\rho\,\vec{u} \dotv\bm{\xi}) = \int\!\d t\!\!\int\!\d V \Pi_\Omega(\vec{x},t)\partial_t(\rho\,\vec{u} \dotv\bm{\xi}) \\
	&= \int\!\d t\!\!\int_\Omega\!\d V \{\partial_t [\Pi_\Omega(\vec{x},t)\rho\,\vec{u}\dotv\bm{\xi}] 
				- \rho\,\vec{u}\dotv\bm{\xi}\,\partial_t \Pi_\Omega(\vec{x},t) \} \\
	&= - \int\!\d t\!\!\int_\Omega\!\d V \rho\,\vec{u}\dotv\bm{\xi}\,\partial_t \Pi_\Omega(\vec{x},t) \;,
\end{split}
\end{equation*}
where the spatial integration range on the right is now arbitrarily large and the $\Pi_\Omega$ is a unit top-hat, (a.k.a. rectangle or boxcar) function with support on $\Omega$.  Restricting to variations with support on $t$ not including the initial and final times, in the last line we have dropped end-point terms from the complete time derivative term $\partial_t[\cdot]$. 

Assuming, as an example, $\Omega$ to be an annular toroid \cite{Dewar_Yoshida_Bhattacharjee_Hudson_2015} we introduce a right-handed, but generally non-orthonormal, curvilinear coordinate system $\{\theta,\zeta,s\}$ where $\theta$ and $\zeta$ are poloidal and toroidal angles, respectively, and $s_\Omega(\vec{x},t)$ is a radial coordinate whose level surfaces are tori, with the surface $s_\Omega = 0$ the inner torus of $\partial\Omega$ and $s = 1$ the outer one. Then $\Pi_\Omega(\vec{x},t) = \Theta(s_\Omega) \Theta(1 - s_\Omega)$, $\Theta(\cdot)$ being the Heaviside step function. Thus we write
\begin{equation*}
\begin{split}
	&\int\!\d t\!\!\int_\Omega\!\d V\partial_t(\rho\,\vec{u} \dotv\bm{\xi})  \\
	&= -\!\!\int\!\d t\!\!\iiint\!\d\theta\d\zeta\d s_\Omega\, \mathcal{J}_\Omega\rho\,\vec{u}\dotv\bm{\xi}\,(\partial_t  s_\Omega)[\delta(s_\Omega) - \delta(s_\Omega - 1)]  \;,
\end{split}
\end{equation*}
where $\mathcal{J}_\Omega = \sqrt{g}$ is the Jacobian of the transformation $\vec{x} \mapsto \{\theta, \zeta, s_\Omega\}$.
The coefficient of the $\delta$ functions, $\partial_t  s_\Omega$, can be evaluated by observing that, as the elements of $\partial\Omega$ are advected with velocity $\vec{v}$, so are the level surfaces of $s_\Omega$ defining $\partial\Omega$. That is,
\begin{equation}
	(\partial_t + \vec{v}\dotv\grad)s_\Omega = 0 \;,
\end{equation}
so $\partial_t s_\Omega = -\vec{v}\dotv\grad s_\Omega \equiv \vec{v}\dotv\esup{s_\Omega}$, where $\esup{s_\Omega}$ is one of the three basis vectors $\{\esup{\theta},\esup{\zeta},\esup{s_\Omega}\} \equiv \{\grad\theta,\grad\zeta,\grad s_\Omega\}$. Thus the final result for the first term of \eqref{eq:deltaSphraw} is 
\begin{align}
	&\int\!\d t\!\!\int_\Omega\!\d V\partial_t(\rho\,\vec{u} \dotv\bm{\xi})  \nonumber\\
	&= \int\!\d t\!\!\iiint\!\d\theta\d\zeta\d s_\Omega\mathcal{J}_\Omega\esup{s_\Omega} \dotv\vec{v} \rho\,\vec{u}\dotv\bm{\xi}\,[\delta(s_\Omega) - \delta(s_\Omega - 1)] \nonumber\\
	&= -\!\!\int\!\d t\!\!\int_{\partial\Omega}\!\d\vec{S} \dotv\vec{v} \rho\,\vec{u}\dotv\bm{\xi} 
	\label{eq:commutedtandint} \;,
\end{align}
where we used the differential geometry identity 
\begin{equation}\label{eq:dS} 
\begin{split}
	\d \vec{S}\equiv \vec{n}\,\d S
	&\equiv \sgn\!(\vec{n}\dotv\grad s_\Omega)\,\esub{\theta}\cross\esub{\zeta}\,\d\theta \d\zeta \\
	&= \sgn\!(\vec{n}\dotv\grad s_\Omega)\,\mathcal{J} \esup{s_\Omega}\,\d\theta \d\zeta \;,
\end{split}
\end{equation}
$\vec{n}$ being the outward normal on $\partial\Omega$ so the sign function $\sgn$ gives $-$ at $s_\Omega = 0$ and $+$ at $s_\Omega = 1$.

Using Gauss' theorem to cast the volume integral over $\divv(\rho\vec{v}\vec{u}\dotv\bm{\xi})$ as a surface integral, and comparing with the result in \eqref{eq:commutedtandint}, we see that the first two terms in $\delta\Sph$, \eqref{eq:deltaSphraw} (which arose from $\rho\vec{u}\dotv\delta\vec{v}$) \emph{cancel}. Thus, inserting $\mathcal{L}_\Omega^{\rm Rx}$ explicitly, we now have
\begin{equation*} 
\begin{split}
	&\delta\Sph = \int\!\d t\!\!\int_{\partial\Omega}\!\d \vec{S}\,\dotv\left\{
	 - \rho\,\bm{\xi}\vec{v}\dotv\vec{u}\phantom{\frac{u^2}{2}}\right. \\
	&+ \rho\,\bm{\xi}\left(\frac{u^2}{2} 
	 	 -\frac{\tau_\Omega }{\gamma - 1}\left[\ln\left(\kappa\frac{p}{\rho^\gamma}\right) 
		  -\gamma\right] \right) \\
	&-\frac{1}{\muSI}\left[(\bm{\xi}\cross\vec{B} + \grad\delta\chi)\cross\!\left(\vec{B} 
	 - \frac{\mu_\Omega}{2} \vec{A} - \XHelmult  \vec{u}\right)\right]\\
	 &+ \bm{\xi}\left[\rho\vec{v}\dotv\vec{u} - \rho\frac{u^2}{2} - \frac{\tau_\Omega\rho}{\gamma - 1} 
	 - \frac{B^2}{2\muSI} \right.\\
	 & \left.+\left. \frac{\mu_\Omega\vec{A}\dotv\vec{B}}{2\muSI} + \frac{\XHelmult\vec{u}\dotv\vec{B}}{\muSI} 
	 + \frac{\tau_\Omega\rho}{\gamma - 1} \ln\left(\kappa\frac{p}{\rho^\gamma}\right)  \right]\right\}
	 \, . 
\end{split}
\end{equation*}
The terms in $\vec{v}\dotv\vec{u}$, $u^2$, and the logarithmic terms cancel, so, expanding and collecting terms,
\begin{equation*}
\begin{split}
	\delta\Sph &= \int\!\d t\!\!\int_{\partial\Omega}\!\d \vec{S}\,\dotv\left\{
	\tau_\Omega\rho\,\bm{\xi} \phantom{\frac{u^2}{2}} \right.\\
	&-\frac{1}{\muSI}\left[(\bm{\xi}\cross\vec{B} + \grad\delta\chi)\cross\!\left(\vec{B} 
	 - \frac{\mu_\Omega}{2} \vec{A} - \XHelmult  \vec{u}\right)\right]\\
	 &+ \left.\bm{\xi}\left[  - \frac{B^2}{2\muSI}
	 +\frac{\mu_\Omega\vec{A}\dotv\vec{B}}{2\muSI} + \frac{\XHelmult\vec{u}\dotv\vec{B}}{\muSI} 
	  \right]\right\} \\ 
	& =  \int\!\d t\!\!\int_{\partial\Omega}\!\d \vec{S}\,\dotv\left\{
	\left(p + \frac{B^2}{2\muSI}\right)\bm{\xi} \phantom{\frac{u^2}{2}} \right.\\
	&-\left. \frac{1}{\muSI}\left[(\grad\delta\chi)\cross\!\left(\vec{B} 
	 - \frac{\mu_\Omega}{2} \vec{A} - \XHelmult  \vec{u}\right)\right] \right\} 	  
	 \, ,
\end{split}
\end{equation*}
where we used the boundary condition $\vec{n}\dotv\vec{B} = 0$, \eqref{eq:tangential}, to eliminate some terms.

\begin{equation}\label{eq:RxActionsurfvariation} 
\begin{split}
	\delta\Sph &= \int\!\d t\!\!\int_{\partial\Omega}\!\d \vec{S}\,\dotv\left\{
	\bm{\xi}\left(p + \frac{B^2}{2\muSI}\right) \right.\\
	&+\left. \frac{1}{\muSI}\left[\delta\chi\curl\!\left(\vec{B} 
	 - \frac{\mu_\Omega}{2} \vec{A} - \XHelmult  \vec{u}\right)\right] \right\} \\
	&= \int\!\d t\!\!\int_{\partial\Omega}\!\!\!\d S\, \vec{n}\,\dotv\bm{\xi}
	\left(p + \frac{B^2}{2\muSI}\right)  
	 \, , 
\end{split}
\end{equation}
where, in the second line, the gauge term in $\delta\chi$ was eliminated using the surface integration by parts identity
\begin{equation}\label{eq:SurfacePartsInteg2} 
	\int_{\partial\Omega} (\grad g) \cross \vec{f} \dotv\d\vec{S} \equiv
	 -\!\!\int_{\partial\Omega} g(\curl\vec{f})\dotv\d\vec{S} \;,
\end{equation}
the Euler--Lagrange equation \eqref{eq:AHamPrEL}  
and the tangential boundary condition \eqref{eq:tangential}. $\Box$ (While we have held the Lagrange multipliers fixed in the calculation,  in principle they also vary during boundary variations to maintain the constancy of their respective constraint functionals. However, the exact constraints do not contribute to the final result \eqref{eq:RxActionsurfvariation} as the Lagrange multipliers have dropped out.)

In Multiregion RxMHD (MRxMHD) each point on $\partial\Omega$ is also on the boundary of a neighboring region, $\Omega'$ say, with unit normal $\vec{n}' = -\vec{n}$. Thus the total action variation from $\bm{\xi}$ localized around such a point is $\delta\Sph + \delta\Sphprime$, giving the standard MRxMHD interface jump condition. \cite[e.g.]{Hudson_etal_2012b, Dewar_Yoshida_Bhattacharjee_Hudson_2015}
\begin{equation}\label{eq:jump}
	\jump{p + \frac{B^2}{2\muSI}} = 0 \;.
\end{equation}

Interestingly, the Galilean-invariant pressure-balance equation \eqref{eq:jump}, which couples neighboring relaxation regions, contains no time derivatives. Yet it is this equation that imparts the inertia of the plasma fluid to interface dynamics, \cite{Dewar_Tuen_Hole_2017} through the effect of the internal RxMHD dynamics within $\Omega$ determining changes in $p$ and $B^2$ at the boundary. 

\section*{Acknowledgments and data statement}
Some of this material is based on work by authors RLD and JWB supported by US National Science Foundation under Grant No. DMS-1440140 while they were in residence at the Mathematical Sciences Research Institute in Berkeley, California during the Fall 2018 semester. We also gratefully acknowledge useful discussions with John Finn, Zensho Yoshida, Philip Morrison, Robert MacKay and Darryl Holm. The research of ZQ was supported by the Australian Research Council under grant DP170102606, and that of JWB was supported by the Los Alamos National Laboratory LDRD program under project number 20180756PRD4. Also RLD, ZQ, and JWB acknowledge travel support from the Simons Foundation/SFARI (560651, AB). 

Data sharing is not applicable to this article as no new data were created or analyzed in this study. More mathematical steps in the equations are given as Supplementary Material in an extended version available online at the same doi as this paper.

\section*{References and footnotes}

\bibliography{RLDBibDeskPapers_MRXMHD}

\end{document}